%% file: ms.tex
\input{definitions}

\documentclass{emulateapj}
\usepackage{natbib,psfig,amsmath}

\shorttitle{The {\it HST}/ACS Coma Cluster Survey. II.}
\shortauthors{Hammer et al.}

\begin{document}
\title{The {\it HST}/ACS Coma Cluster Survey. II. Data Description and Source Catalogs\altaffilmark{1}}
\author{Derek Hammer\altaffilmark{2,3},
Gijs Verdoes Kleijn\altaffilmark{4},
Carlos Hoyos\altaffilmark{5},
Mark den Brok\altaffilmark{4},
Marc Balcells\altaffilmark{6,7},
Henry C. Ferguson\altaffilmark{8},
Paul Goudfrooij\altaffilmark{8},
David Carter\altaffilmark{9},
Rafael Guzm\'an\altaffilmark{10},
Reynier F. Peletier\altaffilmark{4},
Russell J. Smith\altaffilmark{11},
Alister W. Graham\altaffilmark{12},
Neil Trentham\altaffilmark{13},
Eric Peng\altaffilmark{14,15},
Thomas H. Puzia\altaffilmark{16},
John R. Lucey\altaffilmark{11},
Shardha Jogee\altaffilmark{17},
Alfonso L. Aguerri\altaffilmark{6},
Dan Batcheldor\altaffilmark{18},
Terry J. Bridges\altaffilmark{19},
Kristin Chiboucas\altaffilmark{20},
Jonathan I. Davies\altaffilmark{21},
Carlos del Burgo\altaffilmark{22,23},
Peter Erwin\altaffilmark{24,25},
Ann Hornschemeier\altaffilmark{3},
Michael J. Hudson\altaffilmark{26},
Avon Huxor\altaffilmark{27},
Leigh Jenkins\altaffilmark{3},
Arna Karick\altaffilmark{9},
Habib Khosroshahi\altaffilmark{28},
Ehsan Kourkchi\altaffilmark{28},
Yutaka Komiyama\altaffilmark{29},
Jennifer Lotz\altaffilmark{30},
Ronald O. Marzke\altaffilmark{31},
Irina Marinova\altaffilmark{17},
Ana Matkovic\altaffilmark{8},
David Merritt\altaffilmark{18},
Bryan W. Miller\altaffilmark{32},
Neal A. Miller\altaffilmark{33,34},
Bahram Mobasher\altaffilmark{35},
Mustapha Mouhcine\altaffilmark{9},
Sadanori Okamura\altaffilmark{36},
Sue Percival\altaffilmark{9},
Steven Phillipps\altaffilmark{26},
Bianca M. Poggianti\altaffilmark{37},
James Price\altaffilmark{26},
Ray M. Sharples\altaffilmark{11},
R. Brent Tully\altaffilmark{38},
Edwin Valentijn\altaffilmark{4}}

\altaffiltext{1}{Based on observations with the NASA/ESA {\it Hubble Space 
Telescope} obtained at the Space Telescope Science Institute, which is 
operated by the association of Universities for Research in Astronomy, 
Inc., under NASA contract NAS 5-26555. These observations are associated with 
program GO10861.}
\altaffiltext{2}{Department of Physics and Astronomy, Johns Hopkins 
University, 3400 North Charles Street, Baltimore, MD 21218, USA.}
\altaffiltext{3}{Laboratory for X-Ray Astrophysics, NASA Goddard Space 
Flight Center, Code 662.0, Greenbelt, MD 20771, USA.}
\altaffiltext{4}{Kapteyn Astronomical Institute, University of Groningen, 
PO Box 800, 9700 AV Groningen, The Netherlands.}
\altaffiltext{5}{School of Physics and Astronomy, The University of Nottingham,
University Park, Nottingham, NG7 2RD, UK.}
\altaffiltext{6}{Instituto de Astrof\'isica de Canarias, 
38200 La Laguna, Tenerife, Spain.}
\altaffiltext{7}{Isaac Newton Group of Telescopes, Apartado 321, 38700 Santa Cruz de La Palma, Spain.}
\altaffiltext{8}{Space Telescope Science Institute, 3700 San Martin Drive, 
Baltimore, MD 21218, USA.}
\altaffiltext{9}{Astrophysics Research Institute, Liverpool John Moores 
University, Twelve Quays House, Egerton Wharf, Birkenhead CH41 1LD, UK.}
\altaffiltext{10}{Department of Astronomy, University of Florida, PO Box 
112055, Gainesville, FL 32611, USA.}
\altaffiltext{11}{Department of Physics, University of Durham, South Road,
Durham DH1 3LE, UK.}
\altaffiltext{12}{Centre for Astrophysics and Supercomputing, Swinburne 
University of Technology, Hawthorn, VIC 3122, Australia.}
\altaffiltext{13}{Institute of Astronomy, Madingley Road, Cambridge CB3 0HA, UK.}
\altaffiltext{14}{Department of Astronomy, Peking University, Beijing 100871, China.}
\altaffiltext{15}{Kavli Institute for Astronomy and Astrophysics, Peking University, Beijing 100871, China.}
\altaffiltext{16}{Plaskett Fellow, Herzberg Institute of 
Astrophysics, National Research Council of Canada, 5071 West Saanich Road, 
Victoria, BC V9E 2E7, Canada.}
\altaffiltext{17}{Department of Astronomy, University of Texas at Austin, 
1 University Station C1400, Austin, TX 78712, USA.}
\altaffiltext{18}{Department of Physics, Rochester Institute of Technology, 
85 Lomb Memorial Drive, Rochester, NY 14623, USA.}
\altaffiltext{19}{Department of Physics, Engineering Physics and Astronomy, 
Queen's University, Kingston, Ontario K7L 3N6, Canada.}
\altaffiltext{20}{Gemini Observatory, 670 N.~A'ohoku Pl, Hilo, HI 96720}
\altaffiltext{21}{School of Physics and Astronomy, Cardiff University, The 
Parade, Cardiff CF24 3YB, UK.}
\altaffiltext{22}{UNINOVA-CA3, Campus da Caparica, Quinta da Torre, Monte de Caparics 2825-149, Caparica, Portugal.}
\altaffiltext{23}{School of Cosmic Physics, Dublin Institute for Advanced Studies, Dublin 2., Ireland.}
\altaffiltext{24}{Max-Planck-Institut f\"ur Extraterrestrische Physik,
Giessenbachstrasse, D-85748 Garching, Germany.}
\altaffiltext{25}{Universit\"{a}tssternwarte,
Scheinerstrasse 1, 81679 M\"{u}nchen, Germany.}
\altaffiltext{26}{Department of Physics and Astronomy, University of
Waterloo, 200 University Avenue West, Waterloo, Ontario N2L 3G1, Canada.}
\altaffiltext{27}{Astrophysics Group, H.~H.~Wills Physics Laboratory, University of 
Bristol, Tyndall Avenue, Bristol BS8 1TL, UK.}
\altaffiltext{28}{School of Astronomy, Institute for Research in Fundamental Sciences (IPM),
P.O. Box 19395-5531, Tehran, Iran.}
\altaffiltext{29}{Subaru Telescope, National Astronomical Observatory of 
Japan, 650 North A`ohoku Place, Hilo, HI 96720, USA.}
\altaffiltext{30}{Leo Goldberg Fellow, National Optical Astronomy 
Observatory, 950 North Cherry Avenue, Tucson, AZ 85719, USA.}
\altaffiltext{31}{Department of Physics and Astronomy, San Francisco State 
University, San Francisco, CA 94132-4163, USA.}
\altaffiltext{32}{Gemini Observatory, Casilla 603, La Serena, Chile.}
\altaffiltext{33}{Department of Astronomy, University of Maryland,
College Park, MD, 20742-2421, USA}
\altaffiltext{34}{Jansky Fellow of the National Radio Astronomy Observatory. 
The National Radio Astronomy Observatory is a facility of the National 
Science Foundation operated under cooperative agreement by Associated 
Universities, Inc.}
\altaffiltext{35}{Department of Physics and Astronomy, University of California,
Riverside, CA, 92521, USA}
\altaffiltext{36}{Department of Astronomy, University of Tokyo, 7-3-1 Hongo, 
Bunkyo, Tokyo 113-0033, Japan.}
\altaffiltext{37}{INAF-Osservatorio Astronomico di Padova, Vicolo 
dell'Osservatorio 5, Padova I-35122, Italy.}
\altaffiltext{38}{Institute for Astronomy, University of Hawaii, 2680 Woodlawn
Drive, Honolulu, HI 96822, USA.}
\begin{abstract}
The Coma cluster, Abell 1656, was the target of an {\it HST}-ACS Treasury program designed for deep imaging in the F475W and F814W passbands.
Although our survey was interrupted by the ACS instrument failure in early 2007, the partially completed survey still covers $\sim$50\% of the core high-density region in Coma.
Observations were performed for 25 fields that extend over a wide range of cluster-centric radii ($\sim$1.75 Mpc or 1$^{\circ}$) with a total coverage area of 274 arcmin$^{2}$.
The majority of the fields are located near the core region of Coma (19/25 pointings) with six additional fields in the south west region of the cluster.
In this paper we present reprocessed images and \SExtractor\ source catalogs for our survey fields, including a detailed
description of the methodology used for object detection and photometry, the subtraction of bright galaxies to measure faint underlying
objects, and the use of simulations to assess the photometric accuracy and completeness of our catalogs.
We also use simulations to perform aperture corrections for the \SExtractor\ Kron magnitudes based only on the measured source flux and its half-light radius.
We have performed photometry for $\sim$73,000 unique objects; approximately one-half of our detections are brighter than the 10$\sigma$ point-source detection limit at F814W=25.8 mag (AB).
The slight majority of objects (60\%) are unresolved or only marginally resolved by ACS.
We estimate that Coma members are 5\%-10\% of all source detections, which consist of a large population of unresolved compact sources (primarily globular clusters but also ultra-compact dwarf galaxies)
and a wide variety of extended galaxies from a cD galaxy to dwarf low surface brightness galaxies.
The red sequence of Coma member galaxies has a color-magnitude relation with a constant slope and dispersion over 9 magnitudes (-21$<$M$_{F814W}$$<$-13).
The initial data release for the {\it HST}-ACS Coma Treasury program was made available to the public in 2008 August.
The images and catalogs described in this study relate to our second data release.
\end{abstract}

\section{Introduction}
The Coma cluster has been the subject of numerous surveys from X-ray to radio owing to its richness \citep[Abell Class 2;][]{Abell1989}, proximity ($z\sim0.023$), 
and accessibility at high Galactic latitude ($b\sim88$$^{\circ}$).  Coma member galaxies have been critical to our understanding of
galaxy formation and evolution in dense environments, and provide a local benchmark for comparative studies of galaxy evolution in different environments and at higher redshift
\citep[e.g., the morphology-density relation and the Butcher-Oemler effect;][]{Dressler1980,Butcher1984}.
However, the properties of intrinsically faint objects in the Coma cluster are not yet well characterized compared to other local clusters such as Virgo and Fornax.
The resolution and sensitivity afforded by the {\it Hubble Space Telescope-Advanced Camera for Surveys} \citep[{\it HST}-ACS;][]{Ford1998}
has allowed for detailed studies of faint and compact systems in Coma \citep[][hereafter Paper I]{Carter2008}, such as measuring the structural parameters of 
dwarf galaxies (C.~Hoyos et al.~2010, {\it submitted}, hereafter Paper III), constraining the globular cluster (GC) population (E.~Peng et al.~2010, {\it submitted}, hereafter Paper IV),
studying the nature of compact elliptical galaxies \citep[][hereafter Paper V]{Price2009} and ultra-compact dwarf galaxies (UCDs; K.~Chiboucas et al.~2011, {\it in prep}), 
and establishing membership for faint cluster member galaxies by morphology \citep{Chiboucas2010}.

The ACS Coma Cluster Treasury survey was initiated in {\it HST} Cycle 15 following the success of the ACS cluster surveys
performed in Virgo \citep{Cote2004} and Fornax \citep{Jordan2007}.  In contrast to the Virgo and Fornax surveys that targeted individual early-type galaxies,
the Coma fields were arranged to maximize spatial coverage at the cluster core as well as provide targeted observations at an off-center region of Coma.
The advantage of this observing strategy is that it allows for {\it statistical} measurements of faint cluster members
(e.g., the luminosity function (LF); N.~Trentham et al.~2011, {\it in prep}, hereafter Paper VI) while also probing the effects of the cluster environment across a wide range of cluster-centric distance.
Although the Coma cluster is located at a distance of 100 Mpc (5-6 times more distant than Virgo and Fornax), the spatial resolution afforded by {\it HST}-ACS is similar
to current ground-based observations of Virgo and Fornax ($\sim$50 pc).

The ACS Coma Cluster Treasury survey was awarded 164 orbits (82 fields), although the survey was only 28\% complete when interrupted by the ACS failure in early 2007.
A description of the observing program and image reductions is provided in Paper I.
In this paper, the second in the series, we present \SExtractor\ catalogs for our ACS observations.
The catalogs include astrometric and photometric data for $\sim$73,000 source detections,
as well as basic object classifications (i.e.~extended galaxy or point source).
The majority of detections are background galaxies, although we detect several thousand GCs and 
several hundred galaxies that are likely members of the Coma cluster.
The catalogs do not provide an exhaustive list of extended galaxies in Coma, especially for the faintest low surface brightness (LSB) galaxies, for which 
visual inspection of the images is more successful than automated detection.

This paper is organized as follows: imaging data and calibrations in Section 2,
the methodology for creating source catalogs, including a treatment of bright galaxies in Section 3,
simulations to estimate the completeness and photometric accuracy of the source catalogs in Section 4,
and a description of Coma members in Section 5.
As described in Paper I, we assume that the Coma cluster is located at a distance of 100 Mpc ($z=0.0231$), which corresponds to a distance modulus of 35.00 mag
and an angular scale of 0.463 kpc arcsec$^{-1}$ for H$_{0}$$=71$~km s$^{-1}$Mpc$^{-1}$, $\Omega_{\Lambda}$=0.73, and $\Omega_{M}$=0.27.

\section{Data}
\begin{figure}[t!]
\centerline{\scalebox{0.75}{\rotatebox{-90.0}{\includegraphics*[44,164][546,678]{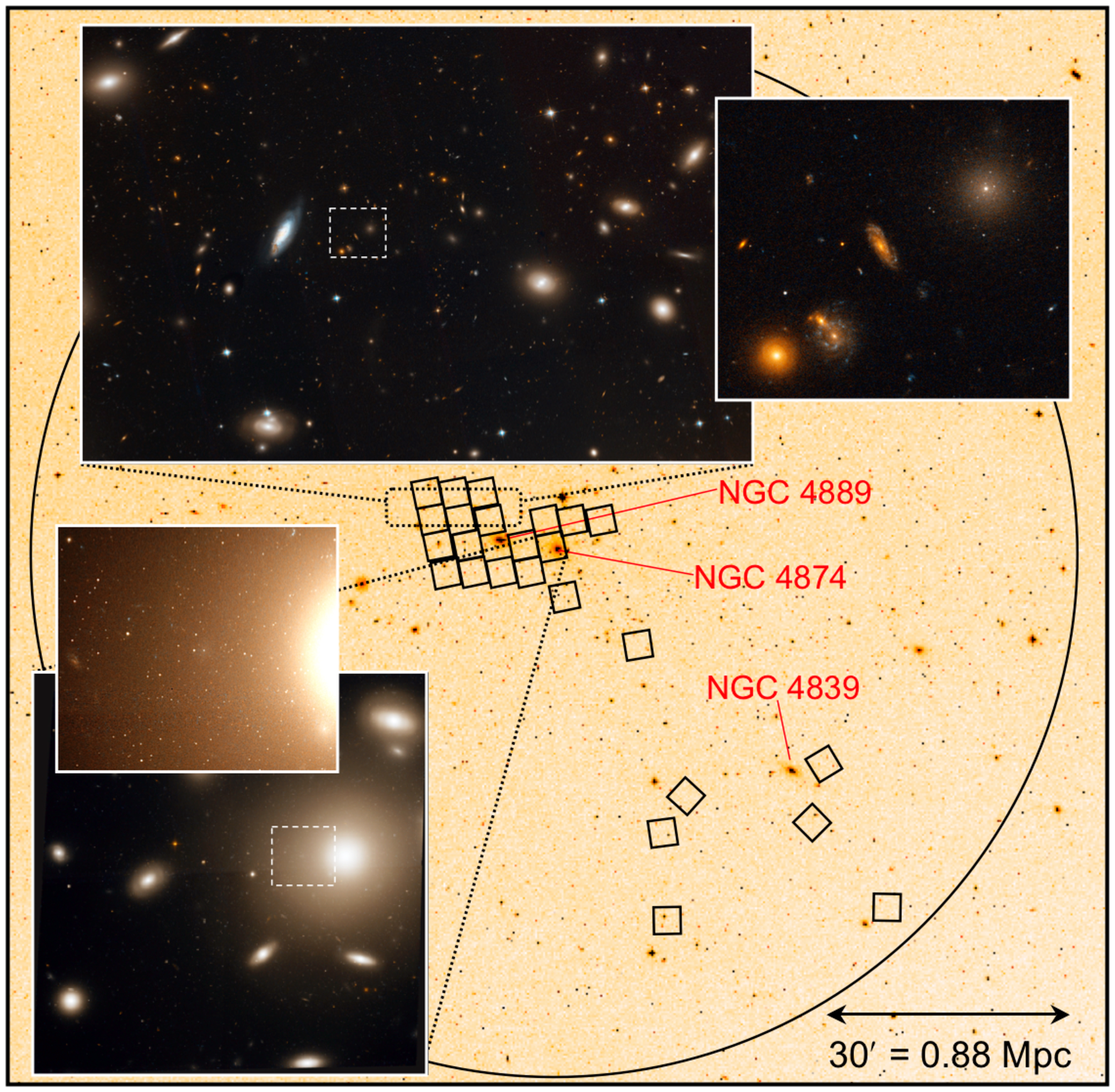}}}}
\caption{\label{field} {\bf View high-resolution images at http://archdev.stsci.edu/pub/hlsp/coma/release2/PaperII.pdf}.
DSS wide-field image of the Coma cluster showing the location of 25 fields observed as part of the {\it HST}-ACS
Coma Cluster Treasury survey (small boxes).
The large circle extends 1$\fdg$1 (1.9 Mpc) from the center of the Coma cluster, or two-thirds the cluster virial radius \citep[r$_{vir}$=2.9 Mpc or 1$\fdg$7;][]{Lokas2003}.
The locations of the three largest galaxies in the Coma cluster (NGC 4889, 4874, 4839) are indicated.
The large inset at top shows a two-color ACS image (F475W=blue and F814W=red) that spans 6 ACS fields;
a subregion of this image (white dashed box) is shown in the small inset, which includes a nucleated dwarf early-type galaxy that is a member of the Coma 
cluster (upper-right corner). The large inset at bottom left shows the entire visit-19 field at the center of the Coma cluster, including the central cD galaxy NGC 4874; a subregion of this
image (located inside the light distribution of NCG 4874; white dashed box) shows the large number of unresolved GCs observed in the galaxy halo.
North is top and east is left.}
\end{figure}

We refer the reader to Paper I for a detailed description of the observing strategy, science objectives,  and the image pipeline for the {\it HST}-ACS Coma Cluster Treasury Survey.
Here we provide only a brief summary of the survey.
The ACS Wide Field Camera, with a field of view of 11.3 arcmin$^{2}$, imaged the Coma cluster in 25 fields with the F475W and F814W filters.
The footprint of our ACS observations is shown in Figure \ref{field}.
The majority of the survey fields (19 of 25) are located within 0.5 Mpc (0$\fdg$3) of the center of Coma\footnote{The center of the Coma cluster is taken as
the location of the central cD galaxy, NGC 4874 ($\alpha$=194$^{\circ}$.89874 and $\delta$ = 27$^{\circ}$.95927).},
mainly covering the regions around the two central galaxies NGC 4874 and NGC 4889 (although NGC 4889 was not observed).
Six additional fields are located between 0.9 and 1.75 Mpc (0$\fdg$5-1$\fdg$0) south-west of the cluster center.
The core and south-west regions of the Coma cluster are sometimes referred to as `Coma-1' and `Coma-3', respectively \citep{Komiyama2002}.

Processed images and initial source catalogs for the ACS Coma Cluster Treasury program were released in
2008 August (Data Release 1)\footnote{Available from Astro-WISE (www.astro-wise.org/projects/COMALS/ACSdata.shtml) and MAST (archive.stsci.edu/prepds/coma/).}.
This paper describes the images and source catalogs provided in Data Release 2.1 (DR2.1), which includes
several enhancements to the initial release such as:
(1) improved alignment between F814W and F475W images and thus aperture-matched color information,
(2) more reliable photometry for extended galaxies,
(3) refinement of the image astrometry,
(4) application of aperture corrections to the \SExtractor\ photometry, and
(5) subtraction of bright galaxies to recover faint underlying objects.

\subsection{Drizzled Images}
The nominal observing sequence for each field consists of four dither positions (`DITHER-LINE' pattern type) each having an exposure
time of 350 s and 640 s for the F814W and F475W filters, respectively. Only two dither positions are available across the 
ACS inter-CCD chip gap resulting in a degraded 3$\arcsec$ band that runs horizontally across the center of every image.
Four fields were observed with less than four dither positions due to the failure of the ACS instrument while the sequence was partially complete.
In Table 1 we list the details for each field including a unique identifier given by the {\it HST} visit number, coordinates, image orientation,
number of dithers, and the integrated exposure time in each band.

The dithered exposures were individually dark- and bias-subtracted, and flat-fielded through the standard {\it HST}-CALACS pipeline software.
The MultiDrizzle software \citep{Koekemoer2002} was used to combine the dither positions and create a final image.
The steps in the MultiDrizzle process include (1) aligning the dithered exposures, (2) identifying cosmic rays using a median filter and `cleaning',
(3) weighting pixels from the individual dithers by the inverse variance of the background/instrumental noise,
and (4) mapping the weighted exposures to an output grid with a pixel size of 0$\farcs$05 ({\it pixfrac}=0.8 and {\it scale} = 1.0).

The identification of cosmic rays by the MultiDrizzle software was inefficient in regions of the image not covered by the full dither pattern.
The {\it lacosmic} routine \citep{vanDokkum2001} was therefore used to assist with cosmic ray identification within the ACS chip gap 
and also across the full image for fields with less than four dither positions. A side effect of the {\it lacosmic} routine is that pixels associated with
extreme surface brightness gradients in the image may be erroneously assigned large (negative or positive) values. Although this affects
only a small number of pixels, we have corrected these pixels in the DR2 images by replacing their values with the local median.
Owing to erratic behavior, we do not apply {\it lacosmic} to regions at the image edge not covered by the full dither pattern,
thus the majority of remaining cosmic rays found in DR2 images are located in these regions.


\subsection{Relative Astrometry}
A slight spatial offset between F475W and F814W images ($\sim$0.3 pixels) was discovered soon after the initial data release.
This offset has a non-negligible effect on the aperture-matched color measurements of unresolved or marginally resolved objects.
Images were realigned by measuring the residual shift between the F475W and F814W frames using compact sources as a reference, 
and then re-drizzling the F475W band to match the original F814W image.
For three fields (visits 3, 10, and 59), it was necessary to re-drizzle the images in both filters to a common pixel position.
The DR2 images are now aligned to within $\sim$0.05 pixels along each axis.

\subsection{Absolute Astrometry}
A two-step process was used to calculate the astrometry for each ACS image, starting with the solution calculated by MultiDrizzle and then performing a
fine adjustment using the \SCAMP\ software \citep[v1.4;][]{Bertin2006}.
The initial spatial registration of images was calculated with MultiDrizzle using reference stars in the {\it HST} Guide Star Catalog.
Astrometry solutions were reliable to only $\sim$1$\arcsec$ (1$\sigma$ rms offset, as compared to the SDSS coordinate system)
owing to the dearth of reference stars in the Coma fields, which are located at high Galactic latitude ($b\sim88$$^{\circ}$).

Second pass solutions for the image astrometry were performed using \SCAMP\ with the goal of aligning ACS images to the SDSS coordinate system.
Both stars and galaxies from SDSS DR6 were selected as reference objects, which increased the ACS and SDSS matched sample to 18-36 objects for each field.
We excluded saturated stars, ACS objects that were blended with nearby sources, or objects located in regions of the ACS image with reduced exposure time.
Astrometry solutions were calculated for each field using a range of values for the \SCAMP\ input parameters {\tt CROSSID\_RADIUS}, {\tt ASTRCLIP\_NSIGMA}, and {\tt FWHM\_THRESHOLDS}.
We forced \SCAMP\ to solve the ACS astrometry {\it without} using higher order distortion terms ({\tt DISTORT\_DEGREES} = 1) as these values are well constrained by the {\it HST}-ACS pipeline \citep{Meurer2003}.
We chose the \SCAMP\ configuration that resulted in both the best reduced chi-squared fit (between 0.9$\simlt$$\chi$$^{2}$$\simlt$1.2) and the smallest final offset between ACS and SDSS reference objects.

This procedure was performed in the F814W band and the astrometry solutions were applied to images in both filters.
The ACS images are now aligned to the SDSS system with a 1$\sigma$ rms spread of 0$\farcs$1-0$\farcs$2 for each coordinate,
which is consistent with the precision of the SDSS astrometry for extended objects \citep{Pier2003}.


\subsection{RMS Maps}
We created rms maps for every image that are used to establish the detection threshold for source identification and to derive photometric errors.
The rms maps were constructed from the inverse-variance maps produced by MultiDrizzle, which are stored in the [WHT] extension of the fits images.
The WHT maps estimate the background and instrumental uncertainties related to the flat-field, dark current, read noise, and the effective exposure time across the image.

We must, however, apply a correction to the rms maps in order to derive accurate photometric errors, i.e.,
the single-pixel rms values cannot be summed in quadrature because MultiDrizzle introduces
a correlation of the noise in neighboring pixels \citep[see][]{Fruchter2002}.
We therefore apply a scaling factor to the rms map to recover its uncorrelated value over length scales larger than a few pixels \citep[e.g.,][]{Casertano2000}.
The corrected rms map is given by:
\begin{equation}
rms =  \frac{1}{\sqrt{WHT \times f}},
\label{rmsmap}
\end{equation}
where $f$ is the scalar correction for correlated noise, commonly referred to as the `variance reduction factor.'

We have measured the variance reduction factor for a subset of our images by selecting a relatively empty 512x512 pixel region inside each image, and then we perform the following steps:
(1) normalization of the pixels by the effective exposure time, (2) masking of real objects identified using a median filter, (3) measurement and subtraction of the background,
and (4) photometry on its autocorrelation image in order to measure the peak pixel flux relative to the total flux, e.g., the correlation factor is $f$=peak/total, where
a value of 1 indicates that the image is uncorrelated at any pixel scale.
We have measured the correction factor for a subset of images in both filters, recovering an average value of $f$=0.77 $\pm$ 0.08 (1$\sigma$ rms). The average value
is applied to all WHT maps as the correction factor depends primarily on the choice of MultiDrizzle parameters {\it pixfrac} and {\it scale} (which are constants in our image pipeline).


\subsection{Flag Maps}
Flag maps were created for every image in order to identify pixels with low effective exposure time.
The effective exposure time varies abruptly across each image owing to regions that are only partially covered by the dither pattern.
These regions include the ACS chip gap and the image border, which cover 5\%-10\% of the total image area.
We have added additional flags to identify pixels that may be affected by very bright galaxies (described in Section 3.1) and for pixels in close proximity to the image edge.
We have assigned a single binary-coded flag value to each pixel that indicate the following:
[{\tt 1}] associated with a bright galaxy, [{\tt 2}] located within 32 pixels of the image edge,
[{\tt 4}] an effective exposure time less than 2/3 the total integration time, and
[{\tt 8}] zero effective exposure.  The order was chosen such that higher flags correspond to regions of the image
where object detection and photometry are more likely to have systematic effects.

The main purpose of the flag maps is to identify large regions of each image with similar completeness limits.
We note that cosmic rays cleaned by MultiDrizzle often leave a few pixels with low effective exposure, but
have little impact on the ability to detect underlying sources.  To avoid flagging these pixels,
we applied a 7 pixel median filter to the effective exposure map prior to imposing the low-exposure criteria.
A few heavily cleaned cosmic ray events remain flagged in each map.

\section{Source Catalogs}
The DR2.1 source catalogs were created using the \SExtractor\ software \citep[version 2.5;][]{Bertin1996}.
\SExtractor\ was operated in `dual-image mode' which uses separate images for detection and photometry.
Dual-image mode allows for a straightforward consolidation of multi-band photometry into a single source catalog, and allows for aperture-matched color measurements.
The F814W band was chosen as the detection image for both filters because it offers higher signal-to-noise ratio (S/N) for most sources, and results in less galaxy `shredding' (i.e.~a single source is
deblended into multiple objects) as galaxy structure tends to be less `clumpy' in F814W as compared to the F475W band.
Object detection was performed on a convolved version of the detection image (Gaussian kernel of FWHM=2.5 pixels) to limit the number of spurious detections.
We used the background rms maps (Section 2.4) to weight the detection threshold across the convolved image, which allows for more reliable detections at faint magnitudes.

\SExtractor\ performs source detection using a `connected pixel' algorithm,~i.e.,~a detection is registered when a specified number of 
contiguous pixels satisfies the detection threshold (set by the {\tt DETECT\_MINAREA} and {\tt DETECT\_THRESH} parameters).
It then performs a second pass analysis of source detections to identify and deblend objects that are connected on the sky
(set primarily by the {\tt DEBLEND\_MINCONT} and {\tt DEBLEND\_NTHRESH} parameters).
Values for the detection and deblend parameters were chosen after testing a wide range of values
and verifying by visual inspection that obvious sources were detected while minimizing the number of spurious detections.
We used a two-fold approach for estimating the background: a {\tt LOCAL} background estimate (performed inside a 64-pixel thick 
annulus around each object) is taken for the majority of sources, and a {\tt GLOBAL} background estimate with a large mesh ({\tt BACK\_SIZE}=256)
is used for extended galaxies with half-light radii larger than 1$\arcsec$. The {\tt LOCAL} background estimate avoids smoothing over 
local fluctuations in the background, while the {\tt GLOBAL} background calculation prevents an overestimate of the background owing 
to flux in the wings of bright extended galaxies (primarily Coma member galaxies).
The full set of SExtractor parameters adopted for this study are presented in the Appendix.

Two sets of detection flags are provided for each object: the internal \SExtractor\ flags and the exposure/location flags described in Section 2.5.
The latter flag values are calculated by using the bitwise inclusive OR operation across pixels in the flag map that satisfied the detection threshold.
The resulting flag values are given by the {\tt IMAFLAGS\_ISO} field in the source catalogs, and the total number of flagged pixels is given by {\tt NIMAFLAGS\_ISO}.
A description of these flags is provided in the Appendix.

\SExtractor\ photometry was performed using a variety of apertures, such as
(1) isophotal apertures which measure the flux only in pixels that satisfy the detection threshold (i.e., {\tt MAG\_ISO} magnitudes, which are recommended for color measurements),
(2) adjustable elliptical apertures that are scaled according to the isophotal light profile (e.g., {\tt MAG\_AUTO} Kron apertures and {\tt MAG\_PETRO} Petrosian apertures), and
(3) a set of nine fixed circular apertures with radii extending between 0$\farcs$06 and 6$\farcs$0.
Magnitudes are reported in the instrumental F475W and F814W AB magnitude system.
We use magnitude zeropoints of 26.068 and 25.937 for the F475W and F814W bands, respectively, 
which were calculated separately for each visit (but resulting in identical zeropoints) following the procedure in the ACS Data Handbook (their Section 6.1.1).
\SExtractor\ estimates the photometric errors by adding in quadrature both the background/instrumental errors taken from the rms maps and the Poisson errors for 
the measured source counts.
Although we do not correct the \SExtractor\ magnitudes for Galactic extinction, the catalogs include the Galactic $E(B-V)$ color excess for each source 
as taken from the \cite{Schlegel1998} reddening maps.
The \SExtractor\ photometry is limited at bright magnitudes by saturated pixels which may be identified from the internal \SExtractor\ flags.
The majority of point sources brighter than $F814W$=19 mag ($F475W$=20) have saturated pixels, while extended galaxies may be saturated 
at magnitudes brighter than $F814W$=15 (no extended galaxies have saturated pixels in the $F475W$ band).

Source catalogs were created separately for each visit.
The ACS fields in the central region of Coma have small overlap (see Figure \ref{field}); thus there are duplicate detections near the edges of these images.
For the purpose of discussing properties of the total source population, we have concatenated source catalogs from the individual
visits using a 0$\farcs$6 search radius to identify duplicate detections (0$\farcs$6 roughly corresponds to the 3$\sigma$ rms uncertainty of our image astrometry);
for each match, we keep the source with a more reliable exposure flag, else we keep the source with the highest S/N.
The duplicate detections ($\sim$1500 object pairs), unfortunately, do not provide a reliable check of the \SExtractor\ photometry as 95\%
of pairs have edge and/or low-exposure flags, and the remaining objects occupy only a small magnitude range at $F814W$$\simgt$25.5 mag.

\begin{figure}[t!]
\plotone{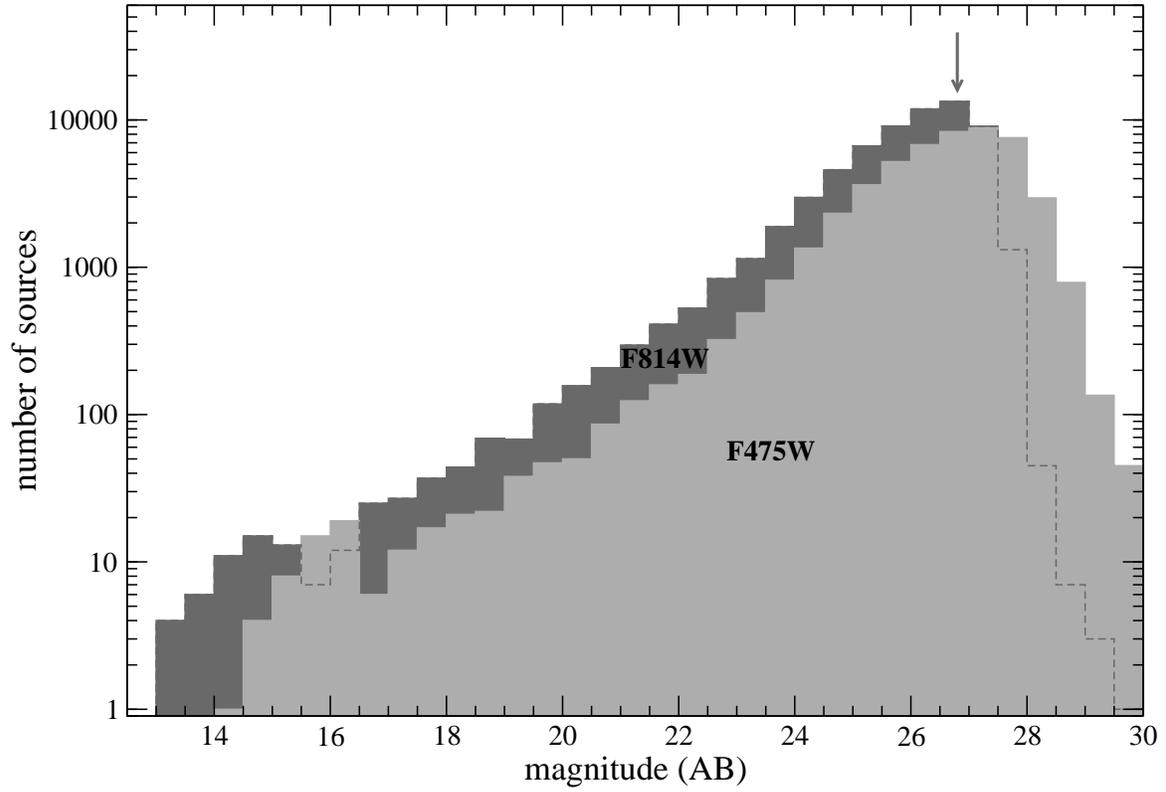}
\caption{\label{mag_histo} Magnitude histogram for catalog sources in both the F814W (dark shaded) and F475W (light shaded) bands.
The F814W magnitudes are taken from the \SExtractor\ Kron photometry and F475W magnitudes are measured within the same F814W Kron apertures.
We do not include F475W sources with S/N$<$3 to avoid unreliable measurements.
The arrow indicates the 80\% completeness limit for point-sources detected in the F814W band as determined by our simulations (Section 4.1).}
\end{figure}

\begin{figure}[t!]
\centerline{\scalebox{0.7}{\includegraphics*[38,192][518,578]{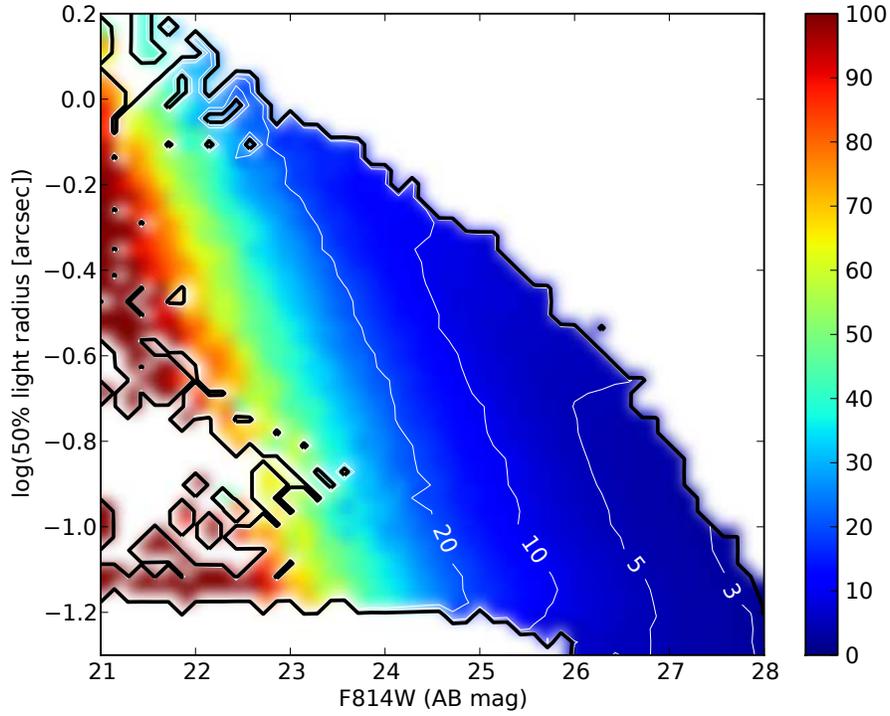}}}
\caption{\label{SN_PLOT} Average signal-to-noise ratio (S/N) for catalog sources in the magnitude-size parameter space:
size is taken as the logarithm of the \SExtractor\ half-light radius in arcseconds, magnitudes are the {\tt MAG\_AUTO} Kron magnitudes,
and S/N is given for the Kron photometry.
The colors correspond to S/N values as defined in the color bar, and white regions of the diagram are void of source detections.
The thin white lines show the contours for average S/N values of 3,5,10, and 20.
Point sources extend roughly horizontal across the bottom of the diagram at log sizes smaller than $\sim$-0.9.
}
\end{figure}

After removing duplicate detections, the catalog consists of $\sim$73,000 unique objects with a magnitude distribution as shown in Figure \ref{mag_histo}.
The raw number counts peak at F814W$\approx$27 and F475W$\approx$27.5 mag, and fall off rapidly at fainter magnitudes.
The average S/N for the \SExtractor\ detections is provided in Figure \ref{SN_PLOT}, shown as contours on a magnitude-size diagram.
Object size is taken as the \SExtractor\ half-light radius in the F814W band ({\tt FLUX\_RADIUS\_3}), which is the size of the circular aperture that encloses 
50\% of the flux in the Kron aperture.  The S/N spans a wide range at a given magnitude owing to intrinsic differences between LSB galaxies detected 
in the foreground Coma cluster and the background galaxies detected at higher surface brightness.
We estimate that $\sim$5\%-10\% of all detections are located inside the Coma cluster (discussed in Section 5.0).
Point sources extend horizontally along the bottom of Figure \ref{SN_PLOT}, which have a 10$\sigma$ (5$\sigma$) point-source detection limit at F814W$\approx$25.8 (26.8) mag,
respectively.

\subsection{Bright Galaxy Subtraction}
We have subtracted the light distribution from bright galaxies in DR2 images in order to improve object detection and photometry for faint underlying sources (e.g., dwarf galaxies, GCs).
The candidates for light subtraction are the 31 Coma member galaxies that have magnitudes brighter than $F814W$=15.0.
We modeled their light distribution using the software package GALPHOT \citep{Franx1989} within the Astro-WISE system \citep[][]{Valentijn2007}.
GALPHOT was chosen because it is a well-defined standard for isophote fitting \citep[e.g.,][]{Jorgensen1995}
and has provided reliable fits for galaxies in {\it HST} images \citep[e.g.,][]{Sikkema2007,Balcells2007}.
GALPHOT uses harmonic fitting for elliptical and higher-order isophotal light models which has the advantage of not requiring 
an initial guess (e.g.,~as compared to the IRAF task ELLIPSE);  note that we fit bright galaxies in visit-19 using both GALPHOT and ELLIPSE 
and visually confirmed that only negligible differences exist between their residual images.

Objects located inside the light distribution of the target galaxy were masked prior to performing our fitting procedure and fits were performed inside image cutouts of each galaxy.
We selected the optimal light model by visual inspection of the residual images, and truncated the best-fitting model at the location where it is indistinguishable from the background noise.
The best-fit light model was subtracted and the cutout was inserted back into the original image.
We could not obtain a satisfactory solution for three bright galaxy candidates:
one spiral galaxy was too irregular to be modeled with elliptical isophotes, and the other two galaxies were overlapping thus preventing a good solution.
Basic properties of the remaining 28 galaxies are listed in Table 2.

In order to perform reliable source detection within regions of galaxy subtraction, we have corrected the rms maps to account for photon noise from the bright galaxy and fitting errors.
Specifically, we added the following term $\sigma_{\rm GSub}^2$ to the variance map:
\begin{equation}
\sigma_{\rm GSub}^{2} = 1.05 \times \sigma_{\rm poisson}^{2},
\end{equation}
where $\sigma_{\rm poisson}$ is the Poisson noise taken from the light distribution model, and we have added an additional 5\% error to account for an imperfect fit \citep[e.g.,][]{delBurgo2008}.
We ran \SExtractor\ on the galaxy-subtracted images (with the updated rms maps) using a configuration that is otherwise identical to that used for the original images.

\begin{figure}[t!]
\centerline{\scalebox{0.8}{\rotatebox{0.}{\includegraphics*[36,307][576,512]{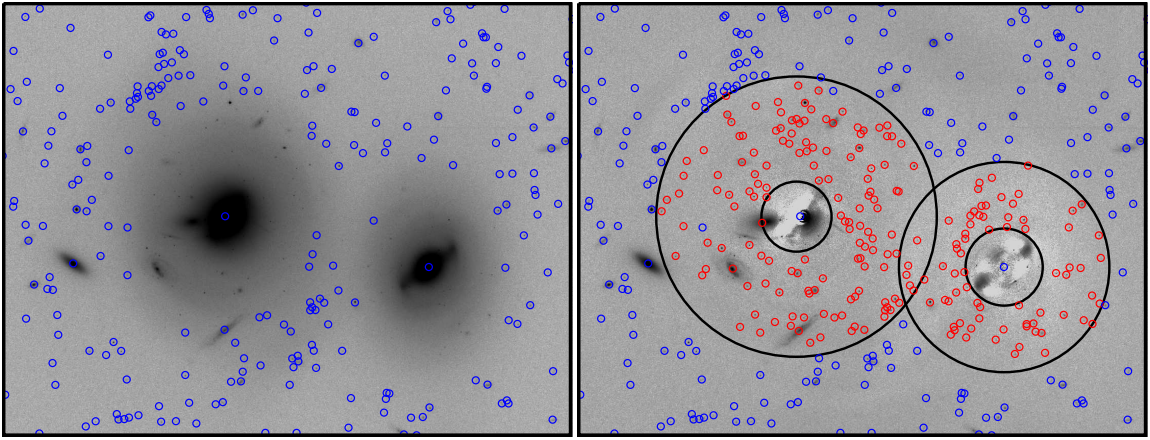}}}}
\caption{\label{Fig:subtracted_galaxies} Example of galaxy subtraction performed for two nearby galaxies in visit-15
(the left panel shows the original image and the right panel shows the residual image after galaxy subtraction).
Blue circles show objects that were detected in the original image, and red circles show objects detected after performing galaxy subtraction.
The concentric circles indicate the annuli used for merging the original and new sources as described in Section 3.1; the inner/outer radius that define the annulus for each subtracted galaxy are listed in Table 2.
Galaxy subtraction was performed by modeling the light distribution using the {\it GALPHOT} software wrapped inside Astro-WISE.
This residual image has the highest noise among all subtracted galaxies owing to the overlapping light profiles.}
\end{figure}

New source detections were merged with the original source catalog using the following procedure.
We defined a circular annulus for each subtracted galaxy with an inner and an outer radius ($r_{\rm min}$, $r_{\rm max}$) as listed in Table 2.
We added to our final source catalog all sources from the galaxy-subtracted catalog that are located in the annulus.
For duplicate detections in the annulus, we adopt photometry from the galaxy-subtracted analysis.
Objects located within the inner radius were removed as large systematic residuals prevent the reliable detection of sources.
At locations beyond the outer radius, we retain the original measurements as both methods include virtually the same real sources.
Figure~\ref{Fig:subtracted_galaxies} shows a pair of overlapping galaxies both before and after performing galaxy subtraction, including the location of original and new source detections.
As a result of subtracting bright galaxies, we have recovered 3291 additional sources that were detected out to 45$\arcsec$ from the centers of the bright galaxies.

\subsection{Object Classification}
In the following section, we present an initial attempt to classify objects, real or otherwise, in the DR2.1 source catalogs.
Object classifications are stored in the catalog {\tt FLAGS\_OBJ} parameter.  A thorough description of these
classifications is given below, and a summary of the flag values is also provided in the Appendix.
We also discuss spurious sources in our catalogs owing to inaccurate galaxy separation
and the chance alignment of random noise.

\subsubsection{Extended Galaxies and Point Sources}
The \SExtractor\ {\tt CLASS\_STAR} parameter, or stellarity index, is a dimensionless value that classifies objects as extended or point-like based on a neural network analysis
that compares the object scale and the image PSF.  The parameter varies between 0 and 1, such that extended objects have values 
near 0 and point-like objects are closer to 1; less reliable classifications are assigned mid-range values.
The galaxy/point source classification is improved by also considering the 50\% light radius
measured in the F814W band. Specifically, we classify as extended galaxies those objects with {\tt CLASS\_STAR}$<$0.5 and 
{\tt FLUX\_RADIUS\_3}$>$2.5 pixels (2.5 pix=0$\farcs$125, or 61 pc at the distance of Coma).
These objects are assigned {\tt FLAGS\_OBJ}=0. Objects that do not meet this criteria are considered point sources and are assigned {\tt FLAGS\_OBJ}=1.
From visual inspection we estimate that this classification scheme is $\sim$95\% reliable for objects brighter than $F814W$=24.0 mag.
The {\tt CLASS\_STAR} parameter is not reliable at fainter magnitudes, thus we use the 50\% light radius criteria alone to classify these objects.
According to this classification scheme, the slight majority of all detections are unresolved (60\%).

\begin{figure}[t!]
\centerline{\scalebox{1.0}{\includegraphics*[30,80][454,380]{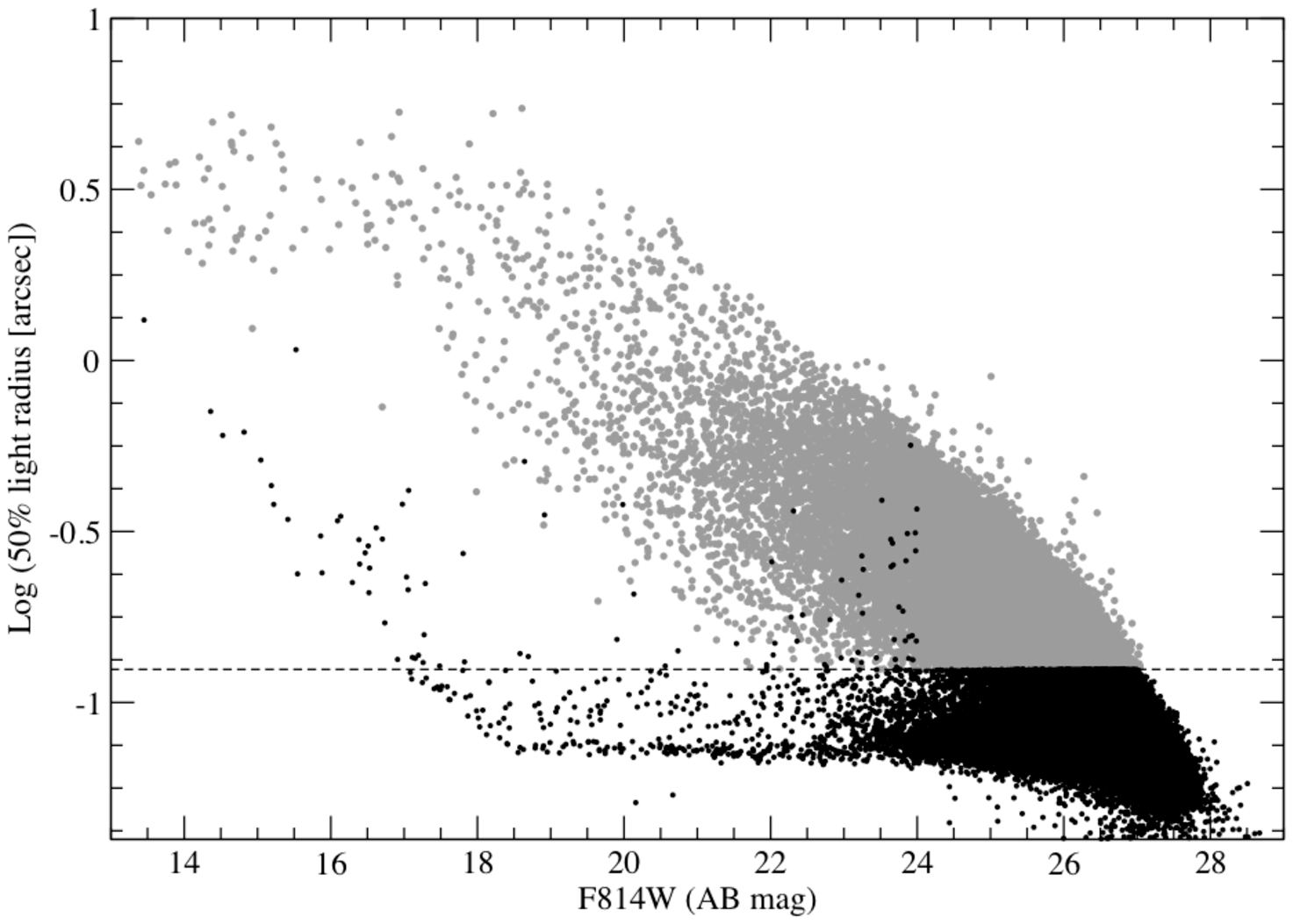}}}
\caption{\label{mag_size} Magnitude-size diagram for catalog sources in the F814W band. Size is taken as the logarithm of the \SExtractor\ half-light
radius in arcseconds, and we use the {\tt MAG\_AUTO} Kron magnitudes.
Gray dots identify extended galaxies and black dots show unresolved sources based on our classification scheme (Section 3.3.1).
The dashed line is the half-light radius used to separate galaxies/point sources at $F814W$$>$24 mag; both the \SExtractor\ {\tt CLASS\_STAR}
parameter and the half-light radius are used at brighter magnitudes. 
Saturated stars are responsible for the extension of point sources at bright magnitudes to larger size.
The unresolved sources that lie above the dashed line between 22$<$F814W$<$24 mag were identified as bright point sources superposed on faint background galaxies.}
\end{figure}

\begin{figure}[t!]
\plotone{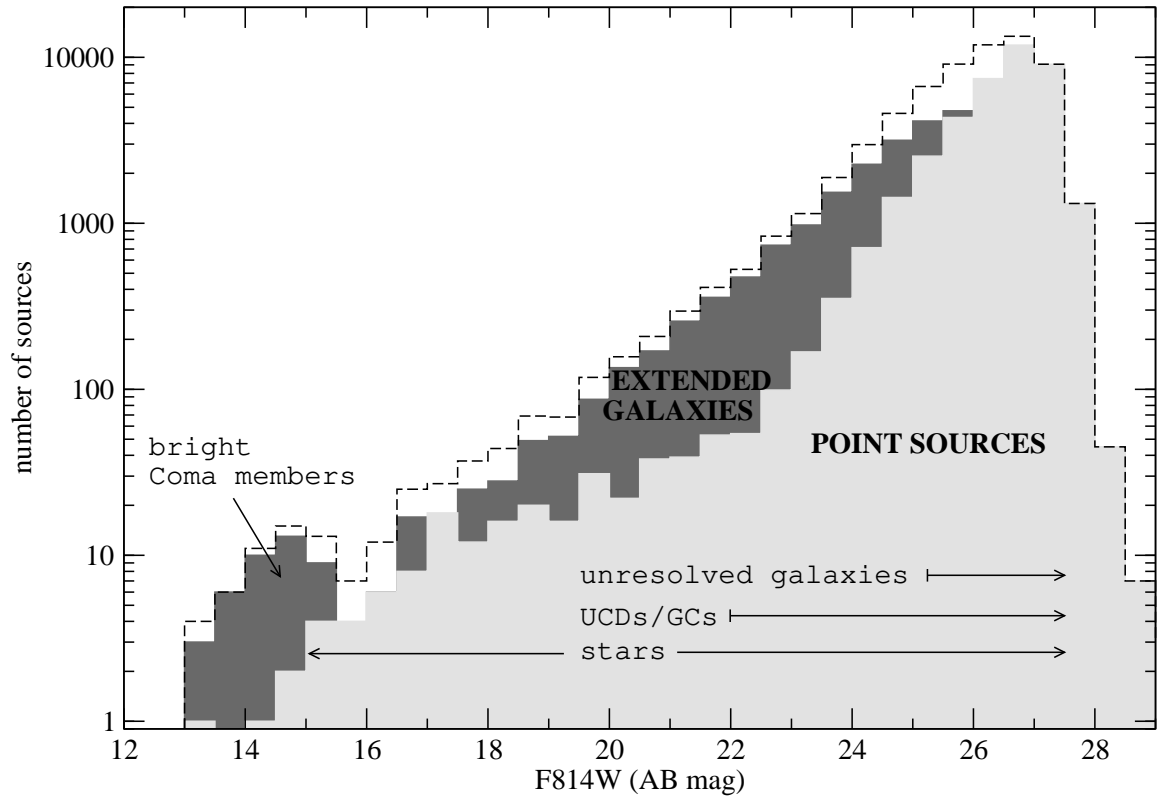}
\caption{\label{maghisto_class} Magnitude histogram for catalog sources in the F814W band.
Separate histograms are shown for objects classified as point sources (light shaded), 
extended galaxies (dark shaded), and the total population (dashed line).
For point sources, we show the magnitude range where we expect contributions from stars, UCDs/GCs, and unresolved galaxies;
UCDs/GCs are likely responsible for the sharp increase in point sources at $F814$=22.5 mag.
Coma members are responsible for the excess number of extended galaxies at bright magnitudes.}
\end{figure}

In Figure \ref{mag_size}, we have plotted catalog sources in a magnitude-size diagram separated as extended galaxies and point sources.
We have visually inspected objects that lie outside their typical magnitude-size parameter space, 
and adjusted {\tt FLAGS\_OBJ} for sources that were obviously misclassified.
Our classification scheme clearly separates extended galaxies and intrinsically compact sources at magnitudes brighter than $F814W$$\approx$25;
at fainter magnitudes, the point source class includes an increasing population of unresolved galaxies.
This can be seen more clearly from the magnitude histogram for extended galaxies and point sources shown in Figure \ref{maghisto_class}.
The histogram for the extended galaxy population flattens at $F814W$=25 mag, or $\sim$1-1.5 mag brighter than the total source population, owing 
to a higher fraction of unresolved galaxies.
The relatively sharp increase in the point source population at magnitudes fainter $F814W$=22.5 may initially reflect
the UCD population in the Coma cluster (K.~Chiboucas et al.~2011, {\it in preparation}), and then the large number of GCs in the cluster with $F814W$$\simgt$24 mag (Paper IV).

\subsubsection{Cosmic Rays}
Although the majority of cosmic rays were removed during assembly of the final images,
a small fraction remain in our catalogs, especially in areas not covered by the full dither pattern.
Cosmic rays typically appear as compact detections that are bright in the F814W image but with no counterpart in the F475W band.
As such, we found that cosmic rays may be identified by selecting candidates with characteristic sizes smaller than the ACS PSF ({\tt FWHM\_IMAGE} $<$ 2.4 pixels),
and then selecting objects with {\tt MAG\_ISO} colors redder than $F475W$-$F814W$$=$3.
Since we perform source detection in the F814W band alone, our catalogs are not well suited for identifying cosmic rays in the F475W images.

Using the above criteria, we identified 18-619 cosmic ray candidates per ACS image at magnitudes brighter than $F814W$=26.5.
We visually inspected the cosmic ray candidates and identified 2-10 real objects per image (primarily stars); 
the remaining cosmic ray candidates are flagged in our source catalogs ({\tt FLAGS\_OBJ} = 2).
We found that cosmic rays are 12\%-25\% of all detections in the border regions not covered by the full dither pattern, but are only $\sim$1\% of objects detected in other regions of the image;
one notable exception is visit-12 where $\sim$50\% (6\%) of objects are cosmic rays in these regions of the image, respectively.

\subsubsection{Image Artifacts}
We have visually identified a small number of detections in our source catalogs that correspond to image anomalies such as diffraction spikes and `ghost' images from bright stars.
For instance, a diffraction spike extends horizontally across the entire image for visit-13 due to a bright star located just outside the field of view.
We have flagged detections that are a direct consequence of image artifacts or real sources whose photometry may be significantly affected by these features ({\tt FLAGS\_OBJ}=3).

\subsubsection{Galaxy Shredding}
We selected values for the \SExtractor\ deblending parameters that minimize the fraction of shredded galaxies, and also result in roughly the same number of shredded objects as
false blended objects for faint field galaxies in ACS images \citep[$\sim$5\%-10\% of detections across $F814W$=23-27 mag are affected by object shredding or blending;][]{Benitez2004}.
The balance between shredded and blended objects does not apply to the regions near bright extended galaxies in the Coma cluster, which tend to include more galaxy shreds;
this is especially the case inside the extended halos of the bright early-type galaxies, where we observe the largest fraction of galaxy shreds.
Such galaxy shreds have little impact on the photometry of the much brighter parent galaxy but result in spurious detections in our catalogs.
We have limited the number of spurious detections by subtracting the light distribution of the 28 brightest galaxies.
This procedure, however, did not remove all galaxy shreds associated with the bright galaxy sample, and many examples remain in our catalogs near galaxies with intermediate magnitudes.

\subsubsection{Random Noise}
Spurious detections may result from the chance alignment of random noise fluctuations.
We have estimated the fraction of such detections by running SExtractor on inverted images, which should give a reasonable estimate provided
that the noise is symmetrical. This analysis applies to empty regions of the image, as it was necessary to mask sources prior to running \SExtractor\ on the inverted image.
We found that $\simlt$1\% of detections are random noise at magnitudes brighter than $F814W$=27.0, but with a rapid increase at fainter magnitudes.
The exception occurs within regions not covered by the full dither pattern for images with fewer than four dither positions (visits 3,12,13,14);
in these cases, $\simgt$10\% of detections are random noise at magnitudes fainter than $F814W$=25.0.

\subsection{Magnitude Comparison with Previous Surveys}
We have compared our catalog to previous optical surveys as a check on the accuracy of the \SExtractor\ photometry.
Several optical photometric surveys have been performed in the Coma cluster \citep[e.g.,][]{Godwin1983,Terlevich2001,Komiyama2002},
but we limit this comparison to the surveys with $I$-band observations that closely match the ACS F814W band,
such as the CFHT deep-field coverage of the central region in Coma \citep{Adami2006CAT} and SDSS DR6 wide-field coverage of Coma \citep{Adelman2008}.
We matched ACS-CFHT galaxies using a 3$\arcsec$ search radius and ACS-SDSS galaxies with a 0$\farcs$6 radius, and limited the comparison
to ACS regions with nominal exposure time and objects with no \SExtractor\ flags.

\begin{figure}[t!]
\plotone{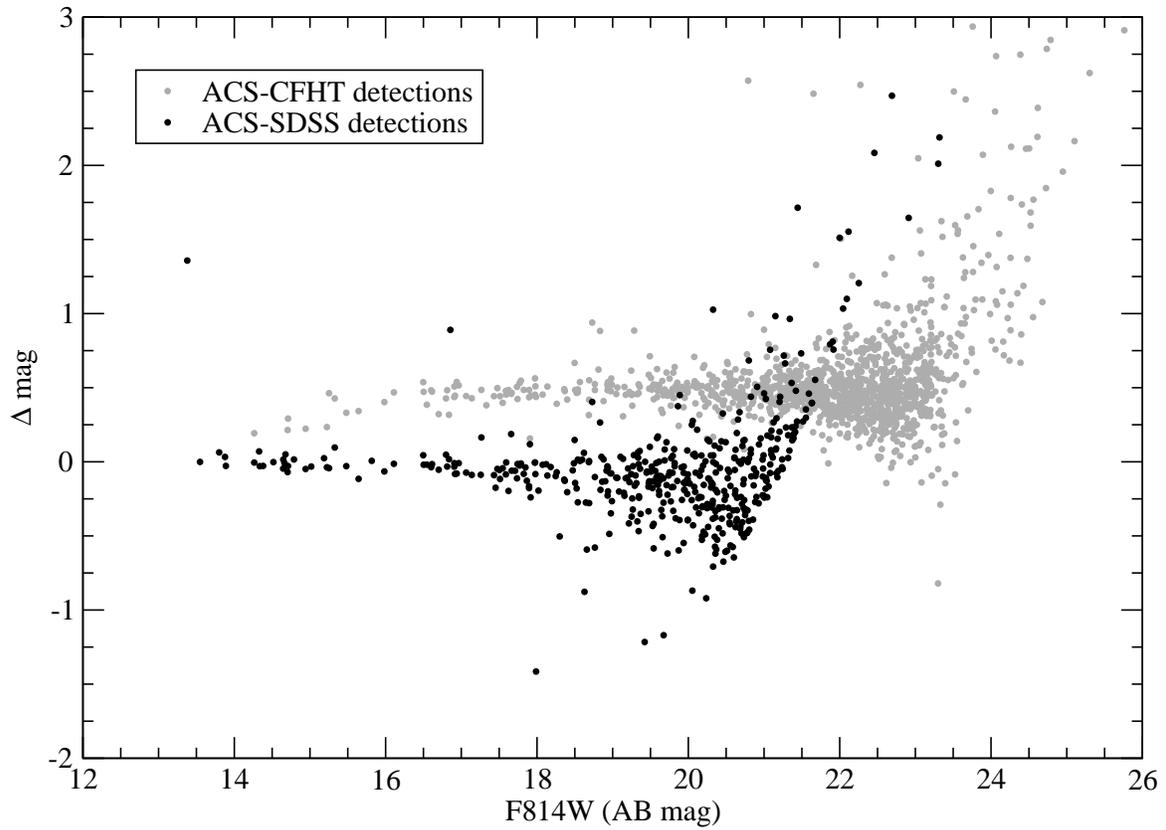}
\caption{\label{mag_compare} Magnitude comparison for galaxies detected in our {\it HST}-ACS observations
and previous optical surveys with $I$-band coverage:
dark circles show galaxies detected in the SDSS $i$-band \citep[instrumental asinh magnitudes;][]{Adelman2008},
and light circles are CFHT $I$-band measurements with the CFH12K camera \citep[instrumental Vega magnitudes;][]{Adami2006CAT}.
We limited this comparison to objects brighter than the SDSS and CFHT completeness
limits. The brightest ACS-SDSS detection is the central CD galaxy NGC 4874, which is too faint in our catalog
owing to its large halo that extends outside the ACS FOV.}
\end{figure}

In Figure \ref{mag_compare}, we show the magnitude offset between our photometry and both CFHT and SDSS as a function of F814W magnitude.
The \cite{Adami2006CAT} CFHT observations are significantly deeper and have better spatial resolution than SDSS,
resulting in less scatter at a given ACS magnitude.
The ACS-CFHT magnitude comparison shows systematic offsets at both bright and faint magnitudes.
From visual inspection, the faint magnitude offset results from both blended objects in the \cite{Adami2006CAT} catalog or shredded galaxies in our catalog.
The offset at bright magnitudes likely results from saturated galaxies in the CFHT observations, which \cite{Adami2006CAT} had replaced with measurements from previous studies.
In contrast, the ACS-SDSS photometry have good agreement at bright magnitudes (F814W$\leq$17), but the systematic offset at fainter magnitudes indicates that
the SDSS flux may be slightly underestimated.
For objects with unreliable ACS measurements, we therefore recommend adopting the SDSS photometry at bright magnitudes (F814W$\leq$17 mag; F814W$_{AB}$$=$SDSS$_{i}$$-$0.01 $\pm$ 0.05)
and CFHT photometry for fainter sources (17$<$F814W$<$22.75 mag; F814W$_{AB}$$=$CFHT$_{I}$$+$0.47 $\pm$ 0.06-0.20 across this magnitude range).

\section{Simulations}
We have tested the reliability of the \SExtractor\ photometry and assessed the completeness limits of our source catalog by injecting synthetic sources
onto the ACS images and re-running our photometry pipeline.
Specifically, we used the \galfit\ software \citep{peng02} to generate a suite of synthetic S\'ersic models that uniformly span a large range in magnitude ($F814W$=20-29, $F475W$=21-30),
effective radius (\Reff\ =0.5-60 pixels), S\'ersic index (\nser=0.8-4.2), and ellipticity ($e$=0.0-0.8).
The models were convolved with a DrizzlyTim/TinyTim PSF \citep[][]{Krist1993} truncated to 63$\times$63 pixels to increase execution speed,
and Poisson noise was added to match the noise characteristics of the data.
For each ACS filter, we randomly inserted a total of $\sim$200,000 models across four fields that cover the cluster core and outskirt regions (visits 1,15,78, and 90), then \SExtractor\ 
was run in a configuration identical to our pipeline as described in Section 2.  The artificial sources were considered detected if \SExtractor\ found a source within 4 pixels ($\sim$2 PSF FWHM) of 
the known position. We refer the reader to Paper III for a more detailed description of these simulations.

In this section, we describe the results of our simulations as it relates to detection efficiency and catalog depth (Section \ref{Sec:Efficiency}), and
the missing-light problem associated with \SExtractor\ photometry (Section \ref{Sec:MissingLight}).

\subsection{Detection Efficiency}
\label{Sec:Efficiency}
It is well known that the \SExtractor\ detection efficiency is a function of object size and magnitude \citep[e.g.,][]{Bershady1998,Cristobal2003,Eliche2006}.
As such, we have measured the detection efficiency separately for models with similar size and magnitude, 
as well as S\'ersic index and ellipticity, and discuss how the detection efficiency depends on these parameters.

\begin{figure}[t!]
\centerline{\scalebox{0.8}{\rotatebox{90}{\includegraphics*[74,138][490,732]{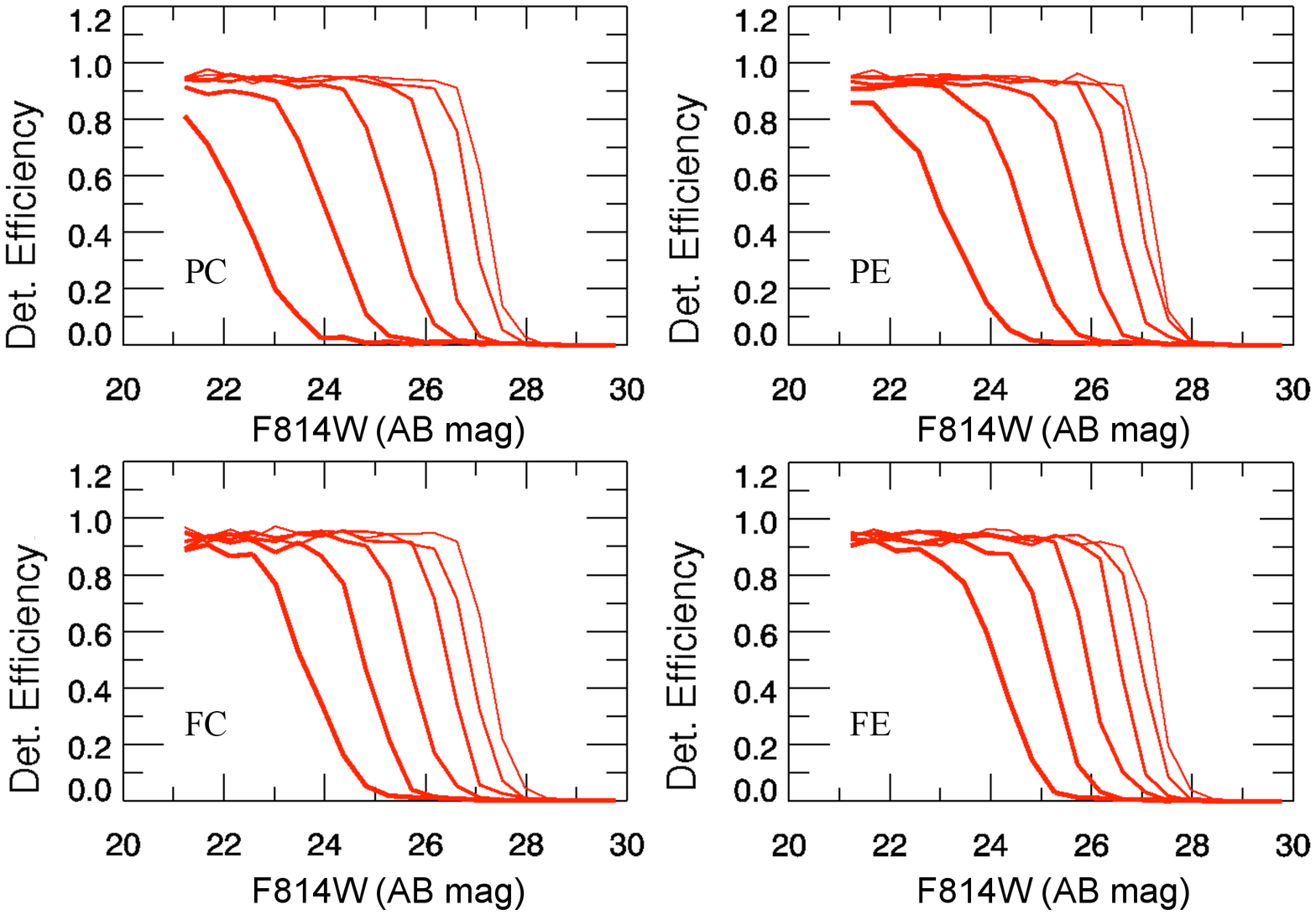}}}}
\caption{\label{Fig:DetEff} Detection efficiency curves as a function of F814W magnitude. The panels show the detection efficiency for
the subset of models (FC,FE,PC,PE) that are separated by S\'ersic index and ellipticity as explained in Section 4.1.
In each panel, we further separate the models into six equally spaced bins in logarithmic \Reff\ (the centers of the \Reff\ bins are identical to the values given in Table 3).
Thin lines represent models with small \Reff, while thick lines are the galaxy models with large \Reff.}
\end{figure}

Figure~\ref{Fig:DetEff} shows the detection efficiency versus magnitude for our models.
We have separated the models into four subsets based on S\'ersic index and ellipticity, which throughout this section 
are referred to as models FC, FE, PC, and PE:
flat (F) and peak (P) models have S\'ersic indices that are smaller/larger than \nser=2.25, respectively; circular (C) and elliptical (E) models have 
ellipticities smaller/larger than $e$=0.4, respectively.
The panels in Figure \ref{Fig:DetEff} show the detection efficiency for each model subset, which we further separate into six equally-spaced bins in logarithmic effective radius that span the range \Reff=0.5-60 pixels.
The left-most curves in each panel trace the largest \Reff\,, which for a given magnitude represent galaxies with the lowest surface brightness.
All efficiency curves show the same pattern:
at the bright end efficiencies are near 100\%, and then fall to zero with a similar slope.
None of the curves reaches 100\% efficiency (maximum rates are $\sim$96\%-98\%) owing to artificial sources that were injected near relatively brighter galaxies.
We note that our \SExtractor\ catalogs reach 100\% efficiency for moderately bright objects upon subtracting the brightest 28 galaxies from our images prior to source detection (Section 3.1).
For a given magnitude, the detection efficiency depends primarily on object size (small \Reff\ gives higher efficiency), followed by S\'ersic index (models with large \nser\ have higher efficiencies),
and to lesser extent on the ellipticity (high-ellipticity models have slightly higher detection rates), i.e.~the detection efficiency is positively correlated with any model parameter that increases the central surface brightness.


In Table 3, we list the expected 80\% completeness limits for our \SExtractor\ catalogs based on our simulations.
The completeness limits are given for the four model subgroups (FC, FE, PC, and PE) separated into the six logarithmically-spaced bins in \Reff\ (F814W results are shown in Columns 2-7, and F475W in Columns 8-13).
The 80\% completeness limits for point sources occurs at $F475W$=27.8 and $F814W$=26.8 mag for images with the nominal four dither positions (corresponding to an average S/N$=$5).
The F814W 80\% completeness limit is consistent with the peak in the magnitude histogram shown earlier in Figure \ref{mag_histo}.

\subsection{Corrections to SExtractor Kron Photometry}
\label{Sec:MissingLight}
Next we have tested the accuracy of the \SExtractor\ photometry by comparing its flux measurements to the magnitudes of the artificial objects.
The \SExtractor\ Kron photometry \citep{Kron1980} is known to capture a varying fraction of the total light,
which is a function of both the S\'ersic profile and S/N,
e.g.,~the \SExtractor\ Kron apertures capture relatively less light for galaxies with high S\'ersic index or low surface brightness \citep[see][]{Graham2005}.
In this section, we derive corrections for the \SExtractor\ Kron magnitudes that provide a better estimate of total magnitudes.
The following analysis is performed for F814W measurements alone, as \SExtractor\ apertures were not defined in the F475W band.

\begin{figure}[t!]
\centerline{\scalebox{0.8}{\includegraphics*[38,220][528,586]{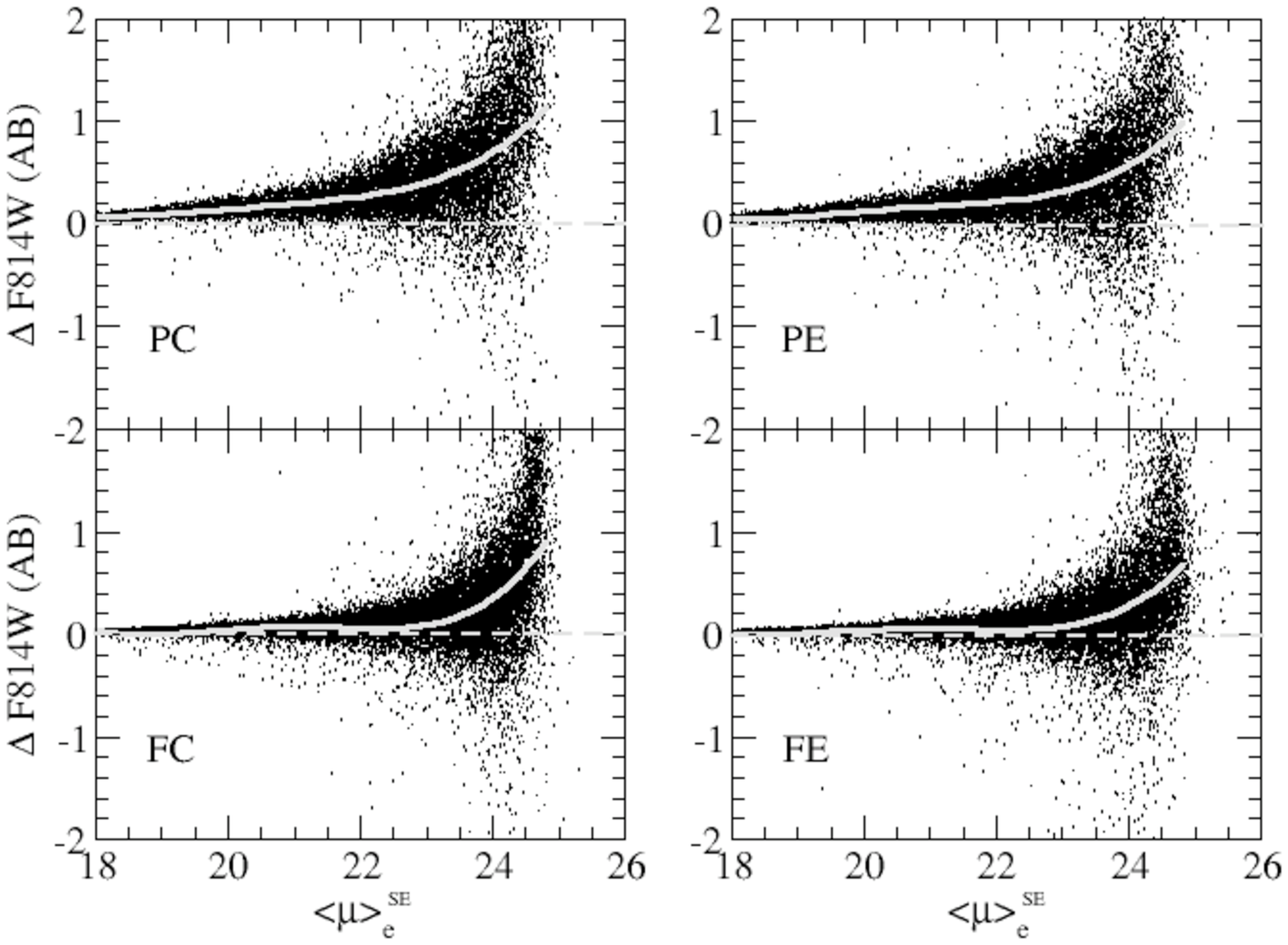}}}
\caption{Magnitude residuals for models detected by \SExtractor\ in the F814W band ({\tt MAG\_AUTO} minus input magnitude) plotted against \meanmueffSE (mag arcsec$^{-2}$ in the F814W band), 
for models injected into the visit-01 image.
The comparison is performed for the subset of models (FC,FE,PC,PE) that are separated based on S\'ersic index and ellipticity as described in Section 4.1.
The solid gray lines show the binned median offset; dashed lines show the zero magnitude offset.}
\label{Fig:RawMagDiff}
\end{figure}

In Figure~\ref{Fig:RawMagDiff} we show the difference between the \SExtractor\ Kron
magnitude and the input magnitude plotted against the \SExtractor\ mean effective surface brightness in the F814W band.
The mean effective surface brightness (mag arcsec$^{-1}$) is defined as

\begin{eqnarray}
\meanmueffSE & \equiv & \mathrm{MAG\_AUTO} + 1.995 + \nonumber \\ 
                   &          &  \mbox{} + 5 \log (\mathrm{PSCALE} \times \mathrm{FLUX\_RADIUS\_3} ),
\label{Eqn:MeanMueEffSE}
\end{eqnarray}

\noindent where \texttt{MAG\_AUTO} and \texttt{FLUX\_RADIUS\_3} are the \SExtractor\ catalog entries for Kron magnitude and the 50\% light radius (in pixels), respectively, 
and \texttt{PSCALE}=0.05 arcsec pixel$^{-1}$ is the plate scale of our ACS images.
We have separated the comparison into four subsets based on S\'ersic index and ellipticity
(models FC, FE, PC, and PE as described in Section \ref{Sec:Efficiency}).
We have trimmed the suite of models used in this analysis to include only those that overlap in 
\Reff\ - magnitude space with galaxies detected in other deep {\it HST} surveys,
e.g.,~the Hubble Ultra Deep Field \citep[][]{hudfpaper}, the Groth Strip Survey \citep[][]{g2dpaper,groth}, and the Galaxy Evolution 
through Morphology and SEDs \citep[GEMS;][]{rix04} surveys.
This was done so that corrections are estimated from models that are representative of the objects potentially detectable in our survey;
the parameter space covered by these models includes the dwarf LSB galaxy population in the Coma cluster \citep[e.g.,][]{Graham2003b}.

The four panels in Figure \ref{Fig:RawMagDiff} show a similar pattern, viz.~photometric errors increase toward fainter surface brightness,
with a systematic bias of the recovered magnitudes toward fainter values.
We also note the following:
\begin{enumerate}

\item{Despite that objects with large S\'ersic indices (``P" models) have a higher detection efficiency,
their \SExtractor\ photometry have both large systematic offset in magnitude and large scatter as compared to other models.}

\item{The systematic magnitude offset for models with low S\'ersic index (``F" models)
is relatively insensitive to surface brightness (i.e.~does not exceed 0.1 mag) for models brighter than \meanmueffSE=23.0.}

\item{The magnitudes of high-ellipticity objects (``E" models) are better recovered than the magnitudes
for objects with low ellipticity (``C" models). The typical magnitude residual of a high-ellipticity model is similar
to the magnitude residual of a low-ellipticity model around 0.5 mag brighter in \meanmueffSE.}

\item{The surface brightness detection limits are the same for all four models (\meanmueffSE=24.75).}

\end{enumerate}

\begin{figure}[t!]
\centerline{\scalebox{0.8}{\includegraphics*[38,220][528,586]{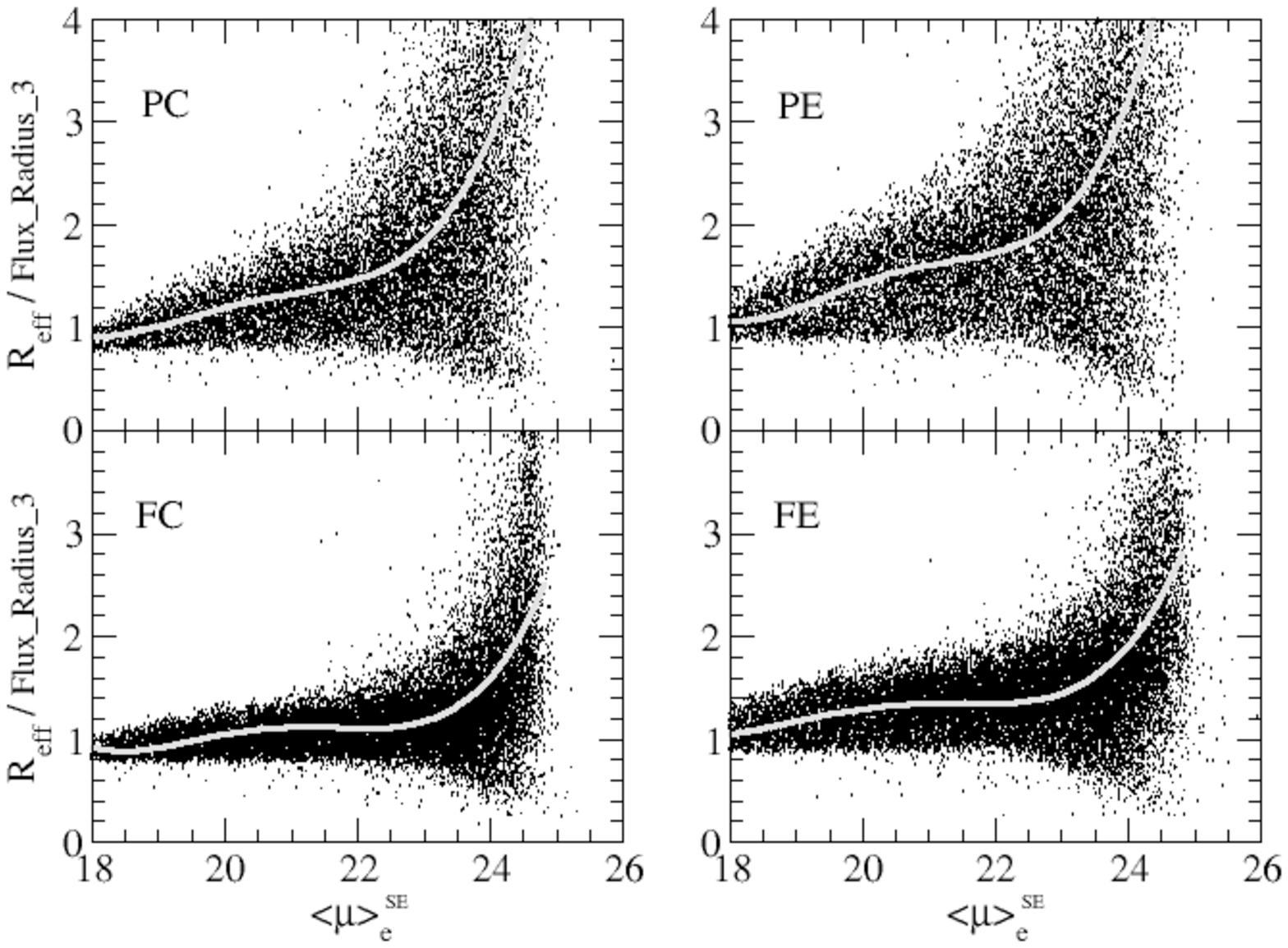}}}
\caption{Ratio of the input effective radius and the output \SExtractor\ half-light radius ({\tt FLUX\_RADIUS\_3}) plotted against \meanmueffSE\ (mag arcsec$^{-2}$ in the F814W band)
for models injected into the visit-01 F814W image. The comparison is performed for the subset of models (FC,FE,PC,PE) that are separated based on S\'ersic index and ellipticity as described in Section 4.1.
Solid gray lines show fifth-order polynomial fits that are used to estimate the true effective radius from the \SExtractor\ half-light radius and surface brightness.}
\label{Fig:ReffDiff}
\end{figure}

In order to correct the \SExtractor\ photometry for missing light we require the true effective radius (\Reff) and the S\'ersic index 
of the galaxy \citep[e.g.,][]{Graham2005}. The \SExtractor\ half-light radius is a proxy for \Reff\ but is often underestimated at faint magnitudes, and also for LSB objects, 
owing to relatively fewer pixels that satisfy the detection threshold.
Instead, we derive an empirical relation between the input effective radius of our models and the output \SExtractor\ half-light radius ({\tt FLUX\_RADIUS\_3}),
which is a function of the S\'ersic profile, ellipticity, and \meanmueffSE. This relationship is shown in Figure \ref{Fig:ReffDiff},
which was fit with a fifth-order polynomial in \meanmueffSE for each of the four model groups.
It is seen that the \texttt{FLUX\_RADIUS\_3} parameter provides a better estimate of the true \Reff\ for 
flat (FE or FC) galaxy models than those with higher S\'ersic indices. The best-fit conversion factor is also more or 
less constant for flat models brighter than \meanmueffSE=23, while it is a steadily growing function of \meanmueffSE\ for 
high-S\'ersic peak models (PC or PE).
This is consistent with the systematic magnitude offsets seen in Figure \ref{Fig:RawMagDiff}.
There are also secondary effects for the galaxy ellipticity, such that \SExtractor\
recovers the true effective radius slightly better for low-ellipticity models.
We conclude that we are able to estimate the true effective radius with good accuracy ($\sim$20\% rms error in the \Reff/{\tt FLUX\_RADIUS\_3} ratio)
based on the \SExtractor\ measurements alone, and thus perform a variable aperture correction to our catalog sources.

\begin{figure}[t!]
\centerline{\scalebox{0.8}{\includegraphics*[38,220][528,586]{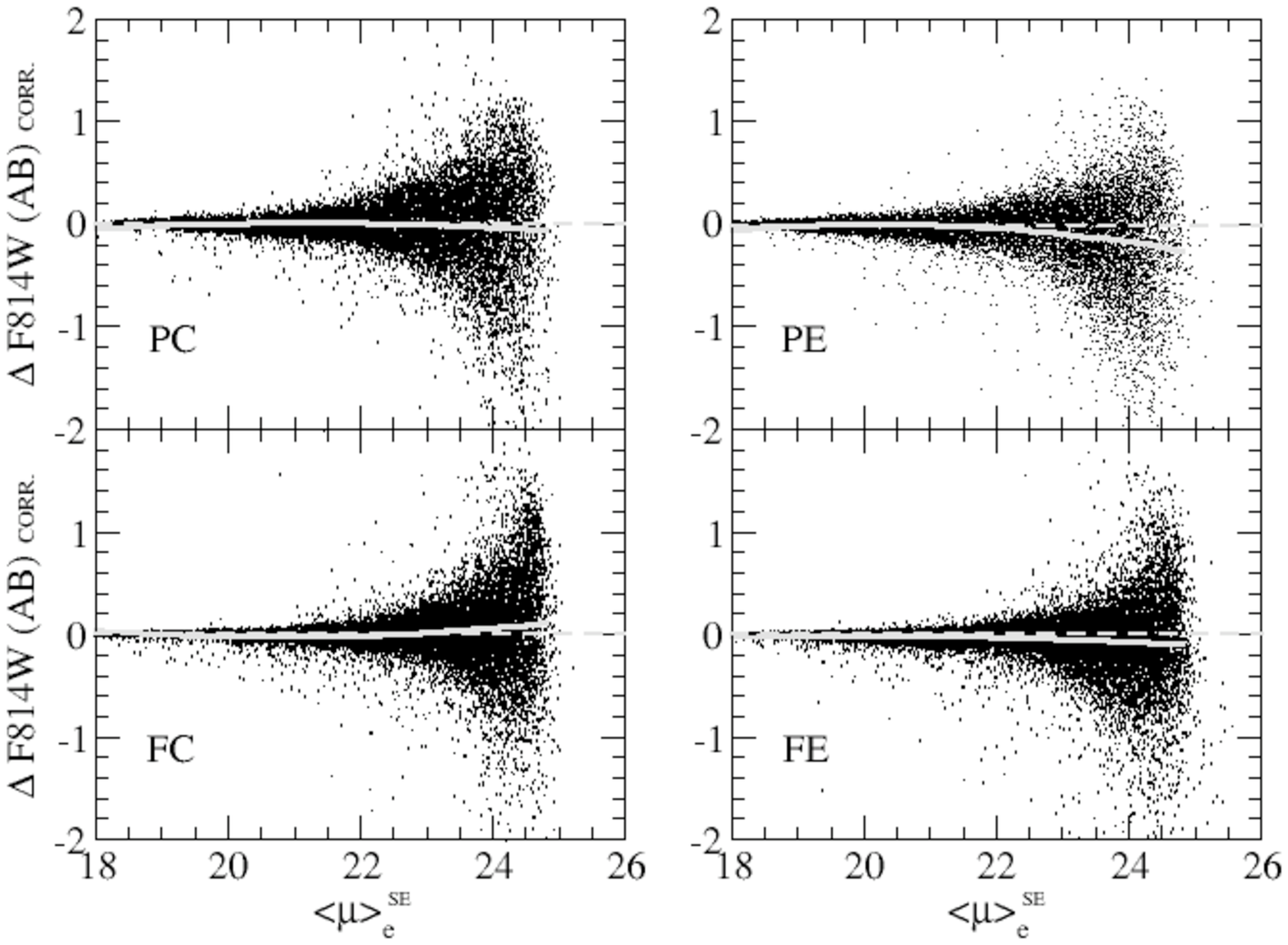}}}
\caption{Same as Figure \ref{Fig:RawMagDiff} except the magnitude residuals are plotted for the aperture-corrected \SExtractor\ Kron photometry.}
\label{Fig:AperCorrMagDiff}
\end{figure}

We have tested the reliability of performing a variable aperture correction on our suite of galaxy models.
The fraction of missing light is calculated based on the ratio of the \SExtractor\ Kron radius and the derived effective radius, and taking the known
S\'ersic index of the models \citep[e.g.,~see Figure 9 in][]{Graham2005}.
The results are shown in Figure~\ref{Fig:AperCorrMagDiff}, which shows the magnitude difference between the
aperture-corrected \SExtractor\ magnitudes and the input magnitude as a function of \meanmueffSE.
The magnitude residuals now follow a well-defined ``trumpet" diagram and are not nearly as 
dependent on galaxy structure/shape as shown in the raw magnitude comparison in Figure \ref{Fig:RawMagDiff}.
There still exists a small systematic offset for all models at low surface brightness (especially for PE models),
for which we will apply a fine correction based on the average residuals measured in bins of surface brightness.
Our aperture correction is reliable for objects with surface brightness brighter than
\meanmueffSE=23.25 in the F814W band. This limit corresponds to our ability to recover the total magnitude
with a 1$\sigma$ rms error of 0.2 mag, but the uncertainties increase rapidly for fainter objects.

We have applied this variable aperture correction to all sources in our \SExtractor\ catalogs.
We do not have prior knowledge of the S\'ersic profile; thus we have performed two aperture corrections
for every object using the following values for the S\'ersic profile:
(1) a flat profile (\nser=1.525) that is representative of the majority of objects in our catalog ({\tt MAG\_AUTO\_CORRA}), and 
(2) a variable S\'ersic profile given by \citet[][]{Graham2003b} that scales with absolute $B$-band magnitude and is applicable to early-type galaxies in the Coma cluster ({\tt MAG\_AUTO\_CORRB}); 
specifically, we use the formula $F475W$=-9.4 log \nser\ + 20.6, assuming a color offset $F475W$=$B$-0.1 for early-type galaxies in Coma and a distance modulus of 35.00 mag.
The first correction is a statistical estimate of the total magnitude and is not meant for accurate measurements of individual galaxies.
The second correction provides a reliable estimate of the total magnitude for individual early-type galaxies in the Coma cluster that are brighter than F814W$=$20;
fainter Coma member galaxies have surface brightness values that are lower than the reliable limit of our aperture correction (\meanmueffSE=23.25).
The errors for the \SExtractor\ corrected magnitudes are taken directly from the measured rms scatter shown in Figure \ref{Fig:AperCorrMagDiff}.
These errors account not only for the counting errors that are measured accurately by \SExtractor, but now include uncertainties related to our ability
to recover the true magnitude. We include an additional 0.05 mag error for both methods to account for the uncertainty of the S\'ersic profile.
The corrections for F475W Kron magnitudes may be estimated indirectly by adding the F475W-F814W color to the corrected F814W Kron magnitude
(assuming the galaxy has a negligible color gradient across the \SExtractor\ Kron aperture).

\section{Objects in the Coma Cluster}
GCs are the majority of Coma members in our catalogs as several thousand are detected at magnitudes fainter than $F814W$$\approx$24 (M$_{F814W}$=-11; Paper IV).
GCs (and most UCDs) are unresolved at the distance of Coma thus their detection efficiency is relatively high as compared to the LSB galaxies that dominate the cluster galaxy population 
at faint magnitudes.
The high density of GCs in the Coma cluster is visible from the inset in Figure 1, which shows the GC population associated with the cD galaxy NGC 4874.
Our catalogs also include several hundred galaxies in the Coma cluster based on:
(1) published redshifts for bright galaxies in our fields \citep[$r$$\simlt$19; e.g.,][]{Colless1996,Mobasher2001,Adelman2008}, and 
deeper spectroscopic redshift surveys performed by team members with MMT-Hectospec ($r$$\simlt$21; R.~Marzke et al., in prep) 
and Keck-LRIS \citep[$r$$\simlt$24;][]{Chiboucas2010}, and
(2) lists of morphology-selected candidates assembled by team members H.~Ferguson and N.~Trentham that provide reliable membership assessments as shown with follow-up spectroscopy
\citep{Chiboucas2010}.

\SExtractor, however, performs poorly for galaxies with very low surface brightness, and hence our catalogs are not exhaustive for the cluster dwarf galaxy population.
From visual inspection, we found that \SExtractor\ detection and photometry is reliable for the majority of dwarf LSB galaxies brighter than F814W$\approx$22 mag.
The exception is for dwarf LSB galaxies located near relatively bright objects, for which \SExtractor\ tends to shred
the galaxy into multiple objects, or does not register a detection.
We have mitigated the fraction of shredded/missing dwarf LSB galaxies by subtracting the light distribution of the 28 brightest cluster member galaxies (Section 3.1).
We have identified $\sim$230 galaxies brighter than F814W=22.5 mag that have both reliable \SExtractor\ photometry and a near 100\% probability of being cluster members. 
Although our morphology assessment suggests that hundreds more cluster LSB galaxies may exist at fainter magnitudes,
these galaxies appear as amorphous structures just visible above the background noise and their detection/photometry is poor regardless of their proximity to other objects.
We will address the nature of these galaxies and their photometry in a future paper.

Cosmological dimming, {\it K}-corrections, and Galactic extinction have important effects
on the photometric measurements of cluster members.
Cosmological $(1+z)^{-4}$ dimming \citep[][]{Tolman1930} affects Coma members such that the observed surface brightness in both ACS bands is
0.1 mag fainter than restframe measurements.
In Table 4, we derived {\it K}-corrections for galaxy SED templates that were redshifted to the distance of Coma and
span a wide range of galaxy morphology and star formation activity \citep[][]{Coleman1980, Kinney1996,Poggianti1997}.
The {\it K}-corrections for Coma members detected in the F475W (F814W) filter span 0.02-0.09 (0.00-0.02) mag for late-type and early-type galaxies, respectively.
These {\it K}-corrections are consistent with values derived with the `{\it K}-corrections calculator'\footnote{http://kcor.sai.msu.ru/} \citep{Chilingarian2010}.
We calculated the Galactic extinction for the SED templates in Table 4 following the formalism in \cite{Cardelli1989} with R$_{V}$=3.1, and
adopting the full range of $E($$B$-$V)$ color excess values for our fields from \cite{Schlegel1998}.
The reddening is relatively insensitive to shape of the galaxy SED at the distance of Coma but varies slightly with color excess,
e.g.,~the Galactic extinction for Coma members is A$_{F475W}$=0.03-0.05 and A$_{F814W}$=0.01-0.03 mag for
color excess values between $E($$B$-$V)$=0.008-0.014, respectively.

\subsection{Color Magnitude Diagram}
\begin{figure}[t!]
\centerline{\scalebox{1.0}{\includegraphics*[40,80][454,380]{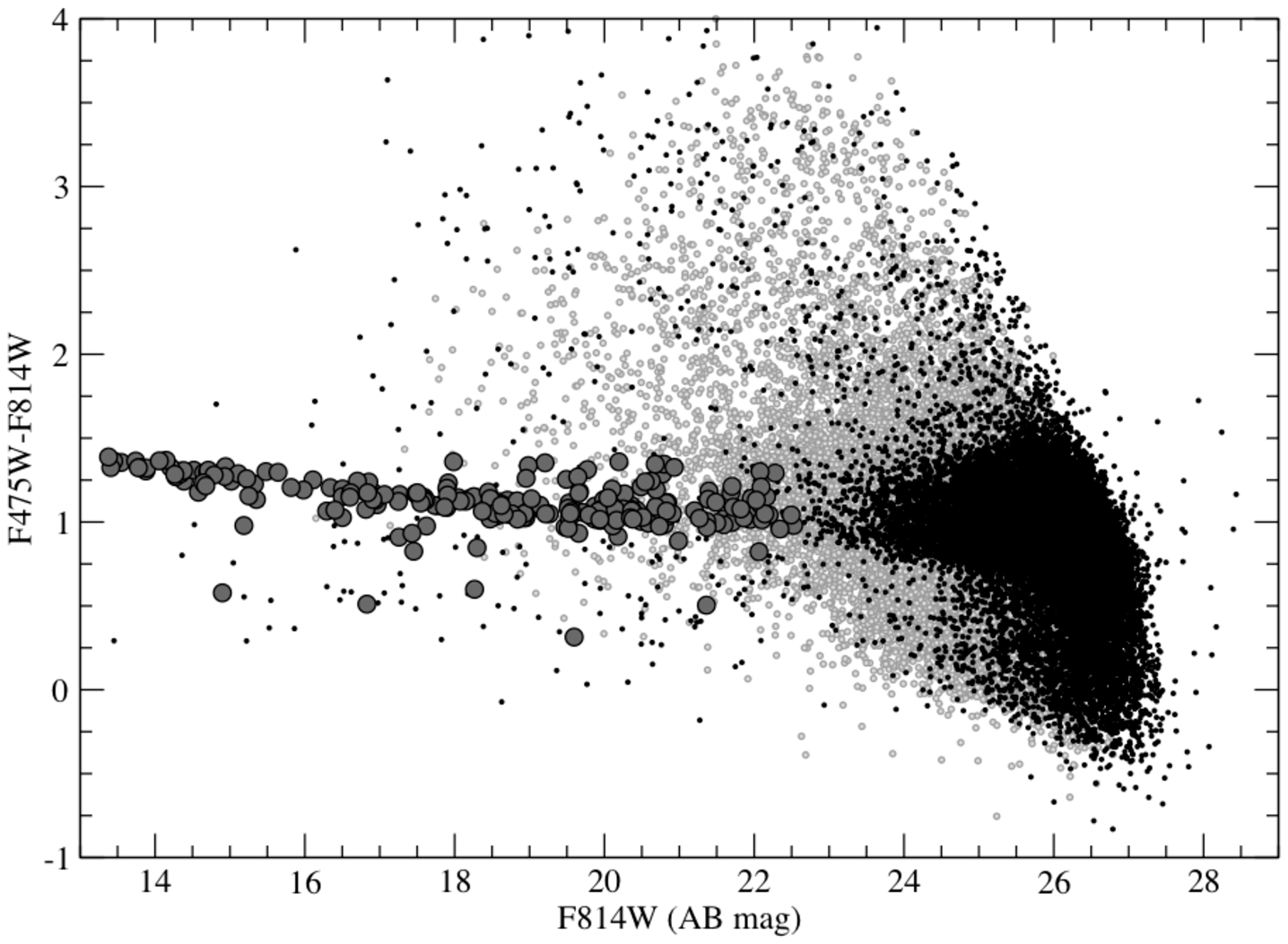}}}
\caption{\label{cmd} CMD for catalog sources separated into extended galaxies (gray dots) and point sources (black dots).
Large filled circles show the location of 233 Coma member galaxies that were identified from spectroscopic redshifts (196/233) and also 
based on morphology (37/233).
Magnitude is taken as the F814W Kron magnitude, and F475W-F814W color is measured using the \SExtractor\ {\tt MAG\_ISO} isophotal photometry.
We do not plot sources with S/N$<$3 in the F475W band.}
\end{figure}

In Figure \ref{cmd}, we show the color-magnitude diagram (CMD) for all point sources and extended galaxies in our catalog, including 
members of the Coma cluster.
Colors are measured inside the \SExtractor\ isophotal apertures as defined in the F814W band.
The GC population is identified as the concentration of point sources at magnitudes fainter than F814W$=$24 mag
with color values near F475W$-$F814W$\approx$1 (Paper IV).
We suspect that unresolved background galaxies are responsible for the extension of point sources to bluer colors at magnitudes fainter than F814W$=$25.
Stars are scattered across other regions of the CMD.

The cluster red sequence (RS) is easily identified as the dense horizontal band of cluster galaxies that stretch between 13$\simlt$F814W$\simlt$22.5 mag.
It is well known that the cluster RS consists primarily of quiescent early-type galaxies, and that massive RS galaxies (as traced by luminosity) have redder
colors owing to their deep potential wells that are able to retain metals for future generations of stars with cooler main sequence turnoffs \citep[e.g.,][]{Bower1992,Kodama1997}.
In contrast, RS dwarf galaxies have bluer colors owing to both feedback mechanisms that remove processed elements \cite[e.g., supernovae winds;][]{Dekel1986} and
more extended star formation histories \citep[e.g.,][]{Smith2008,Sanchez2009}.

\begin{figure}[t!]
\plotone{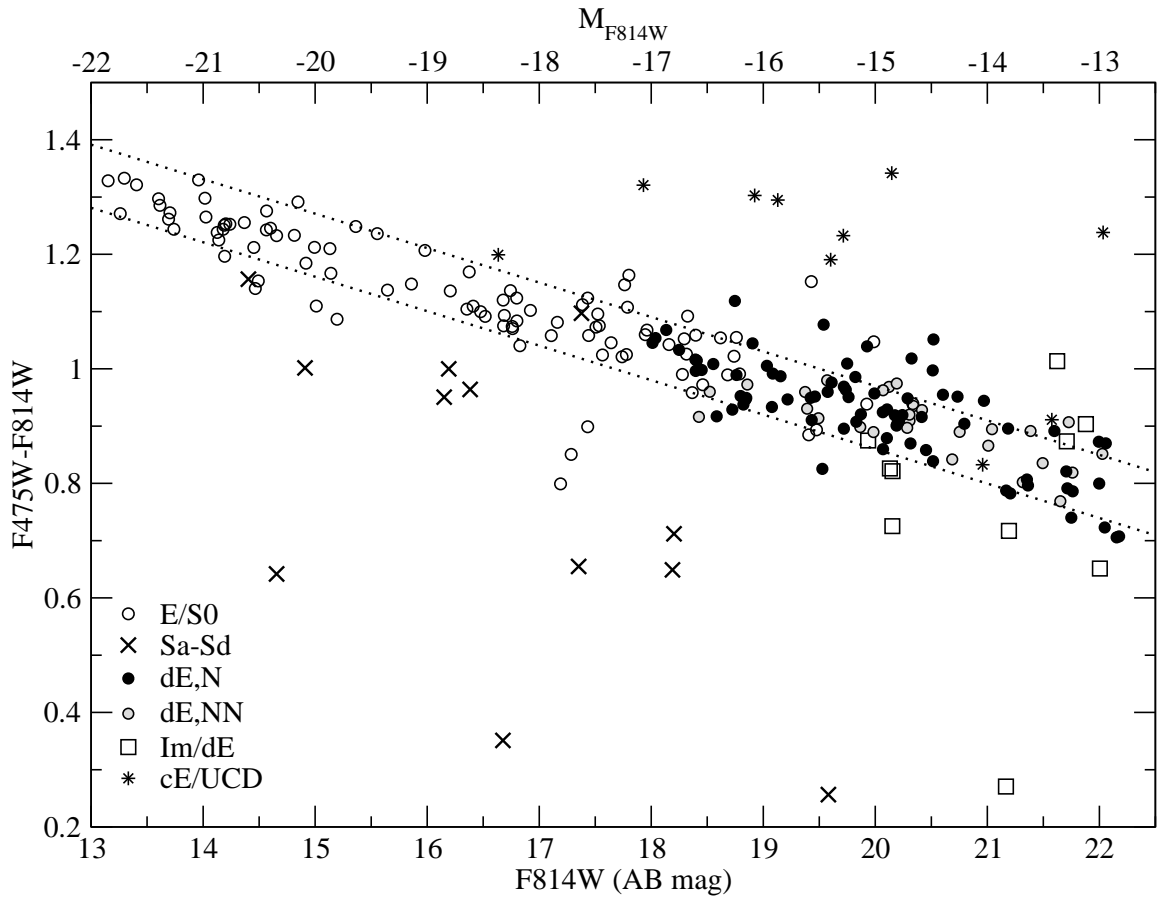}
\caption{\label{cmd_coma} CMD for the Coma member galaxies shown in Figure \ref{cmd}. Symbols separate objects by morphology as
defined in the legend. Magnitudes and colors are corrected for Galactic extinction and {\it K}-corrections, and we use the corrected 
Kron photometry as a measure of the total magnitude.  Colors are measured inside fixed apertures with sizes as defined in Section 5.1.
Dotted lines show the $\pm$1$\sigma$ rms dispersion (0.06 mag) from the fitted relation: F475W-F814W$=$-0.06$\times$F814W + 2.12.}
\end{figure}

Figure \ref{cmd_coma} shows a separate CMD for objects in the Coma cluster.
Magnitudes and colors are adjusted for Galactic reddening and {\it K}-corrections, and we use the Kron magnitudes corrected for missing light (Section 4.2).
Colors were measured inside fixed circular apertures with sizes that were selected to match the typical \Reff\ of cluster RS galaxies:
a 3\farcs0 radius (1.4 kpc) for galaxies brighter than F814W=16 mag, a 2\farcs25 radius (1.0 kpc) for galaxies between 16$<$F814W$<$20 mag, and
a 1\farcs5 radius (0.7 kpc) for fainter galaxies \citep[e.g.,][]{Graham2003,Ferrarese2006}.
Fixed apertures avoid the color bias owing to variable apertures that may not sample the same physical regions in galaxies \citep[][]{Scodeggio2001}.
The slope of the RS appears constant across the full magnitude range, which confirms the results of previous studies \citep[e.g.,][]{Secker1997,LopezCruz2004,Adami2006CAT},
but extends the relation to a few magnitudes fainter.
A robust bisector fit to the full RS gives the relation F475W-F814W$=$-0.060 $\times$ F814W + 2.12 mag, with a 1$\sigma$ dispersion of 0.06 mag.
This relation is not affected by the relatively large errors for the corrected Kron photometry at faint magnitudes, e.g.,
we obtain an identical fit to the RS for galaxies brighter than F814W$=$20 mag.

Interestingly, the RS dispersion remains roughly constant (0.06 mag) when measured in discrete magnitude intervals across the full RS.
This result is in contrast to the relatively large 0.3 mag dispersion in $B$-$R$ color reported for faint ($R$$\sim$22 mag) dwarf galaxies of both the Coma cluster \citep{Adami2009SPEC}
and Perseus cluster \citep{Conselice2003}, which were attributed to either varied star formation histories or effects from dynamical interactions.
Large uncertainties for their color measurements, or contamination from background galaxies, may have obscured an underlying tighter relation with the RS.
Other possible explanations for the low dispersion reported here are:
(1) our redshift coverage and visual classification scheme may preferentially select faint Coma member galaxies that lie near the RS, and
(2) our ACS fields do not cover the regions of the Coma cluster where many faint dwarf galaxies were reported
to be offset from the RS \citep[e.g., west of the cluster center, and in the immediate vicinity of NCG 4889;][]{Adami2009LSB}.
Future studies will provide a more accurate analysis of the RS using magnitudes and colors obtained by fitting the light distribution of each individual Coma member galaxy.

In Figure \ref{cmd_coma}, we have separated Coma members into broad morphology classes based on visual inspection and following the galaxy templates
of \cite{Sandage1984}. This classification scheme is only meant to provide a cursory description of the Coma cluster morphologies and to identify large-scale trends.
Future papers will provide less subjective classifications based on multi-component fits to the light distributions of Coma member galaxies.
As expected, early-type galaxies (E/S0) are the majority of bright RS galaxies and most are located in the central ACS fields,
while spiral galaxies (Sa-Sd) are located below the RS and were detected primarily in the outer ACS fields.
Although we did not consider the SExtractor measurements as part of our classification, our E/S0 and dE galaxy classes can be reproduced with a separation
in surface brightness at \meanmueffSE=22 mag arcsec$^{-1}$ in the F814W band.
Galaxies classified as dEs all have magnitudes fainter than F814W$\approx$18 (M$_{F814W}$=-17),
which corresponds to a B-band magnitude of M$_{B}$$\approx$-16.
The dE galaxies were subdivided into two classes: (1) nucleated dE galaxies (dE,N) have unresolved central nuclei
and are the dominant subclass, especially among bright dE galaxies, and
(2) non-nucleated dE galaxies (dE,NN) lack a visible central nucleus but are otherwise similar in appearance to dE,N galaxies, and are an increasing fraction of total dE population at fainter magnitudes.
Compact galaxies are located primarily above the RS and consist of the seven cE candidates discussed in \cite{Price2009} and another three UCD-like objects that have redshifts.
Ten galaxies have irregular morphology but are conservatively classified as Im/dE as it is difficult to differentiate between these galaxy types at faint magnitudes.

Analysis of the subgroups of early-type galaxies requires more detailed measurements than presented here, but we note that
our classes show no obvious trends in the CMD shown in Figure \ref{cmd_coma}.
The exception, however, relates to the RS outliers at intermediate magnitudes (17.5$\simlt$F814W$\simlt$21) as: (1) several E/S0 and dE,N galaxies are located above the 1$\sigma$ RS but are rare below the RS,
and (2) there is a relative lack of dE,NN galaxies outside the 1$\sigma$ RS.
Interestingly, the majority of these outliers occupy the same region of the CMD as cEs and UCDS, which are thought to originate from the tidally-stripped nuclei of E/S0 and dE,N galaxies \citep[e.g.,][]{Bekki2001,Bekki2003,Price2009}.
It is intriguing to speculate that the outliers above the RS may have experienced at least low-level tidal interactions, or represent a phase in cE/UCD creation,
but such analysis is beyond the scope of this paper and will be explored in future works.

Nearly all galaxies classified as Im/dE (9/10) are located either above or below the 1$\sigma$ RS.
These galaxies may be analogs of the scattered population of faint dwarf LSB galaxies detected in the Coma and the Perseus clusters \citep{Adami2009LSB, Conselice2003}, albeit 
as a much smaller fraction of the total galaxy population than reported in these other studies.
\cite{Adami2006LSB} suggested that tidal stripping may be responsible for similar faint red galaxies in their sample, while 
the galaxies blueward of the 1$\sigma$ RS may be tidal dwarf galaxies formed from the stripped disk material.
Our observations are somewhat consistent with this scenario such that, unlike the bright blue spirals that are found primarily in the outer ACS fields,
all of the Im/dE galaxies (blue or red) were detected in the central ACS fields where interactions are more common.
The nature of these galaxies will be studied in future papers of this series.

A graphical representation of the CMD is presented in Figure \ref{coma_zoo},
which shows two-color postage-stamp images for a subset of our Coma member galaxies.
The color axis is normalized to the RS and color intervals are in standard deviations of the measured dispersion (1$\sigma$=0.06 mag).
This diagram demonstrates the wide range of morphology, star formation activity, and evolution of cluster members detected in our survey, such as:
bright E/S0 galaxies and barred disk galaxies that follow the RS and extend to the dE galaxy population at fainter magnitudes (middle row);
compact galaxies located above the 3$\sigma$ RS (top row) that include cEs at intermediate magnitudes and a UCD at the extreme faint end;
star forming galaxies located below the 3$\sigma$ RS (bottom row);
and more rare galaxies in our sample such as possible interacting galaxies (Row 1, Column 2), an anemic spiral (Row 2, Column 2),
and a faint dwarf irregular galaxy located well below the RS (bottom-right corner).

\section{Summary}
We have improved our image pipeline and created \SExtractor\ catalogs for 25 fields observed for the {\it HST}-ACS Coma Cluster Treasury survey.
The processed images and source catalogs are publicly available for download as part of our second data release
(DR2.1)\footnote{http://www.astro-wise.org/projects/COMALS/ACSdata.shtml}.
The source catalogs include photometry for $\sim$73,000 unique objects that were detected in F814W images, and color measurements for the F475W band.
We performed simulations that indicate our source catalogs are 80\% complete for point sources at $F814W$=26.8 mag;
the simulations were also used to establish aperture corrections to the \SExtractor\ Kron photometry.
The majority of catalog sources are background galaxies, and we estimate that $\sim$5\%-10\% of objects are located inside the Coma cluster.

The majority of Coma members are unresolved GCs,
but also include a wide range of cluster member galaxies such as UCDs, compact elliptical galaxies,
dwarf early-type galaxies (nucleated and non-nucleated), dwarf irregular and spiral galaxies, barred disk galaxies \citep{Marinova2010}, and a cD galaxy (NGC 4874).
Preliminary analysis of the color-magnitude relation indicates that the red sequece of Coma member galaxies has a flat slope that extends to the faintest dE galaxy detections at M$_{F814W}$$=$-13.0 mag (AB),
and has constant dispersion across the full magnitude range (1$\sigma$=0.06 mag).
Early-type galaxies that are offset from the red sequence at magnitudes between -17.5$<$M$_{F814W}$$<$-14 tend to have both redder optical colors than expected and visible nuclear light components.
These relations will be studied in more detail using galaxy properties derived from multi-component fits to the light distribution of Coma member galaxies.
Additional analysis of the Coma member population is ongoing, including but not limited to the following:

\begin{trivlist}

\item [$\bullet$] C.~Hoyos et al.~(2010, {\it submitted}, Paper III) measure the structural parameters (e.g.,~S\'ersic index, effective radius)
for $\sim$50,000 objects selected from the Paper II photometric catalogs. Fits are performed using both GALFIT and GIM2D,
and a detailed comparison of the useful limits for both methods is explored.

\item [$\bullet$] E.~Peng et al.~(2010, {\it submitted}, Paper IV) identify a large population of intracluster globular clusters (IGCs) that are not associated 
with individual galaxies, but fill the core of the Coma cluster and make up $\sim$40\% of its total GC population. The majority of IGCs 
are blue and metal-poor, suggesting they were ejected from the tidal disruption of dwarf galaxies, although 20\% of IGCs may originate from the halos of L$_{*}$ galaxies.

\item [$\bullet$] \citet[][Paper V]{Price2009} investigate a sample of compact galaxies that consists of old intermediate-metallicity
compact elliptical (cE) galaxies, and fainter compact galaxies with properties that reside somewhere between cEs and UCDs.
The measured light profiles, velocity dispersions, and stellar populations are consistent with a formation mechanism
owing to tidal mass-loss from galaxy-galaxy interactions.

\item [$\bullet$] N.~Trentham et al.~(2011, {\it in prep}, Paper VI) measure the galaxy LF to very faint magnitude ($M_{F814W}$=-12).
The cluster LF is measured at different cluster-centric radii and separated by morphological class in order to study the environmental dependence 
of the galaxy population.
Cluster membership is established from both morphology and also using deep redshift coverage with MMT-Hectospec (R.~Marzke et al.~2011, {\it in prep})
and follow-up spectroscopy of ACS sources with Keck-LRIS \citep{Chiboucas2010}.
\end{trivlist}



\acknowledgements
We thank the referee for their helpful comments, Panayiotis Tzanavaris for assistance with the SCAMP software, Antara Basu-Zych
for useful science discussion, Karen Levay for implementing the data release on MAST, and Zolt Levay for constructing the two-color images.
This research and associated EPO program are supported by STScI through grants HST-GO-10861 and HST-E0-10861.35-A, respectively.
Partial support is also provided for the following individuals:
Carter and Karick are supported by UK STFC rolling grant PP/E001149/1;
Erwin is supported by DFG Priority Programme 1177;
Balcells is supported by the Science Ministry of Spain through grants AYA2006-12955 and AYA2009-11137;
Hudson is supported by NSERC;
Guzman is supported by Spanish MICINN under the Consolider-Ingenio 2010 Programme grant CSD2006-00070;
Merritt is supported by grants AST-0807910 (NSF) and NNX07AH15G (NASA).


\clearpage

\clearpage
\bibliographystyle{apj}
\bibliography{ms}
\clearpage


\clearpage

\clearpage

\clearpage

\clearpage

\clearpage

\clearpage

\clearpage

\clearpage

\clearpage

\clearpage

\clearpage

\clearpage

\clearpage

\begin{figure}[t!]
\caption{\label{coma_zoo} {\bf View high-resolution images at http://archdev.stsci.edu/pub/hlsp/coma/release2/PaperII.pdf}.
CMD for a subset of Coma member galaxies shown as two-color image cutouts (F475W=blue and F814W=red).
The {\it y}-axis is the difference in color between galaxies and the cluster red sequence (1$\sigma$=0.06 mag) such that the middle row is the spine of the red sequence.
Coma member detections are not available in the two upper-left panels of the diagram.}
\end{figure}
\clearpage

\include{tab1}
\clearpage
\include{tab2}
\clearpage
\include{tab3}
\clearpage
\include{tab4}
\clearpage

\appendix
\section{Appendix A: SExtractor Configuration and Catalog Description}
We ran \SExtractor\ in dual-image mode using the F814W band as the detection image for both ACS filters.
The \SExtractor\ input parameters are given below in Table \ref{SExtractorParams}.
The \SExtractor\ source catalogs provide over 100 measurements for each object.
In Table \ref{FlagValue}, we describe the three sets of flags given for each object, and Table \ref{CatalogParams} 
defines all other catalog entries.

\begin{table}[h]
\centering
\caption{Parameters for SExtractor version 2.5}
\begin{tabular}{p{4cm} c}
\hline\hline
\tablewidth{0pt}
\tablenum{5}
Parameter & Value       \\ [0.5ex]
\hline
ANALYSIS\_THRESH\dotfill\               &       0.9     \\
BACK\_FILTERSIZE\dotfill\               &       3       \\
BACK\_SIZE\dotfill\                     &       128   (256 for extended galaxies)  \\
BACKPHOTO\_THICK\dotfill\               &       64      \\
BACKPHOTO\_TYPE\dotfill\                &       LOCAL   (GLOBAL for extended galaxies) \\
CLEAN\dotfill\                          &       Y       \\
CLEAN\_PARAM\dotfill\                   &       1.0     \\
DEBLEND\_MINCONT\dotfill\               &       0.03    \\
DEBLEND\_NTHRESH\dotfill\               &       32      \\
DETECT\_MINAREA\dotfill\                &       5       \\
DETECT\_THRESH\dotfill\                 &       0.9     \\
FILTER\dotfill\                         &       Y       \\
FILTER\_NAME\dotfill\                   &       gauss\_2.5\_5x5.conv \\
GAIN\dotfill\                           &       median exposure time (sec) \\
MAG\_ZEROPOINT\dotfill\                 &       26.068 (F475W), 25.937 (F814W) \\
MASK\_TYPE\dotfill\                     &       CORRECT \\
PHOT\_APERTURES\dotfill\                &       2.4,8,20,30,60,90,120,180,240   \\
PHOT\_AUTOPARAMS\dotfill\               &       2.5, 3.5        \\
PHOT\_PETROPARAMS\dotfill\              &       2.0, 3.5        \\
PHOT\_FLUXFRAC\dotfill\                 &       0.2,0.3,0.5,0.8,0.9     \\
PIXEL\_SCALE\dotfill\                   &       0.05    \\
SATUR\_LEVEL\dotfill\                   &       85,000/GAIN     \\
SEEING\_FWHM\dotfill\                   &       0.12    \\
WEIGHT\_GAIN\dotfill\                   &       Y       \\
WEIGHT\_TYPE\dotfill\                   &       MAP\_RMS, MAP\_RMS       \\
WEIGHT\_THRESH\dotfill\                 &       1000000, 1000000     \\ [0.5ex]
\hline
\end{tabular}
\label{SExtractorParams}
\end{table}

\begin{table}[h]
\centering
\caption{Description of Source Catalog Flags}
\begin{tabular}{p{4cm} l}
\hline\hline
\tablewidth{0pt}
\tablenum{6}
Flag Value & Description       \\ [0.5ex]
\hline
\hline \multicolumn{2}{c}{\bf SExtractor Internal Flags ({\tt FLAGS})} \\ \hline
0\dotfill\               &       no flags    \\
1\dotfill\               &       {\tt MAG\_AUTO} photometry may be biased by nearby neighbors ($>$10\% of area is affected)      \\
2\dotfill\               &       object was originally blended with another source     \\
4\dotfill\               &       one or more pixels is saturated   \\
8\dotfill\               &       object is truncated at image boundary   \\
16\dotfill\		 &	 aperture data are incomplete or corrupted \\
\hline \multicolumn{2}{c}{\bf Exposure and Image Location Flags ({\tt IMAFLAGS\_ISO})} \\ \hline
0\dotfill\               &       object pixels have nominal exposure time and location    \\
1\dotfill\               &       object pixel(s) are located within region of subtracted bright galaxy      \\
2\dotfill\               &       object pixel(s) are located within 32 pixels of the image edge     \\
4\dotfill\               &       object pixel(s) have an effective exposure less than two-thirds the nominal exposure   \\
8\dotfill\               &       object pixel(s) have zero effective exposure (e.g.,~object is truncated by image edge)   \\
\hline \multicolumn{2}{c}{\bf Object Flags ({\tt FLAGS\_OBJ})} \\ \hline
0\dotfill\               &       extended galaxy    \\
1\dotfill\               &       point source (e.g., star, GC, UCD, or unresolved background galaxy)    \\
2\dotfill\               &       cosmic ray  \\
3\dotfill\		 &       image artifact \\
\hline
\end{tabular}
\label{FlagValue}
\end{table}
\clearpage

\begin{table}[h]
\centering
\caption{Description of Source Catalog Fields}
\begin{tabular}{p{4cm} l}
\hline\hline
\tablewidth{0pt}
\tablenum{7}
Parameter & Description       \\ [0.5ex]
\hline
COMA\_ID\dotfill\					&	Unique ID assigned to each object (filter+ra+dec).  \\
NUMBER\dotfill\					&	SE sequential value assigned to sources in each catalog (corresponds to value in segmentation map). \\
\hline \multicolumn{2}{c}{\bf Photometric Parameters} \\ \hline
BACKGROUND\dotfill\				&	SE background at centroid position [e-/sec]. \\
BACK\_METHOD\dotfill\				&	SE background method (LOCAL or GLOBAL). \\
FLUX[ERR]\_AUTO\dotfill\			&	SE flux measured within elliptical Kron aperture [e-/sec]. \\
FLUX[ERR]\_ISO\dotfill\				&	SE F475W color flux measured across F814W isophotal pixels [e-/sec]. \\
FLUX\_MAX\dotfill\					&	SE peak flux (above background) measured across isophotal pixels [e-/sec]. \\
FLUX\_RADIUS\_[1-5]\dotfill\			&	SE radius of circular aperture that encloses 20,30,50,80, \& 90\% of FLUX\_AUTO [pix]. \\
KRON\_RADIUS\dotfill\				&	SE Kron radius in units of A\_IMAGE and B\_IMAGE [pix]. \\
MAG[ERR]\_APER\_[1-9]\dotfill\ 		&	SE magnitude measured within fixed circular apertures (0.12,0.4,1,1.5,3,4.5,6,9,12" diam) [AB]. \\
MAG[ERR]\_APER\_[1-9]\_F475\dotfill\	&	SE F475W color magnitude measured in F814W fixed circular apertures [AB]. \\
MAG[ERR]\_AUTO\dotfill\				&	SE magnitude measured within elliptical Kron aperture [AB]. \\
MAG[ERR]\_AUTO\_CORRA\dotfill\		&	Corrected Kron magnitude for a typical galaxy (\nser=1.525) [AB]. \\
MAG[ERR]\_AUTO\_CORRB\dotfill\		&	Corrected Kron magnitude for early-type Coma member (\nser\ from \cite{Graham2003b}) [AB]. \\
MAG[ERR]\_AUTO\_F475\dotfill\		&	SE F475W color magnitude measured in F814W Kron aperture [AB]. \\
MAG[ERR]\_ISO\dotfill\ 				&	SE isophotal magnitude measured across pixels that satisfy detection threshold [AB]. \\
MAG[ERR]\_ISOCOR\dotfill\ 			&	SE corrected isophotal magnitude assuming Gaussian profile [AB]. \\
MAG[ERR]\_ISO\_F475\dotfill\			&	SE F475W color magnitude measured across F814W isophotal pixels [AB]. \\
MAG[ERR]\_PETRO\dotfill\ 			&	SE magnitude measured within elliptical Petrosian aperture [AB]. \\
MU\_MAX\dotfill\					&	SE peak surface brightness (above background) measured across isophotal pixels [mag arcsec$^{-1}$]. \\
MU\_THRESHOLD\dotfill\				&	SE detection threshold in surface brightness above BACKGROUND [mag arcsec$^{-1}$]. \\
PETRO\_RADIUS\dotfill\				& 	SE Petrosian radius in units of A\_IMAGE and B\_IMAGE [pix]. \\
THRESHOLD\dotfill\					&	SE detection threshold above BACKGROUND [e-/sec]. \\
\hline \multicolumn{2}{c}{\bf Geometric Parameters} \\ \hline
A[ERR]\_IMAGE\dotfill\				&	SE RMS profile of isophotal pixels along major axis [pix]. \\
AWIN[ERR]\_IMAGE	\dotfill\			&	SE windowed (Gaussian) RMS profile of isophotal pixels along major axis [pix]. \\
B[ERR]\_IMAGE\dotfill\				&	SE RMS profile of isophotal pixels along minor axis [pix]. \\
BWIN[ERR]\_IMAGE	\dotfill\			&	SE windowed (Gaussian) RMS profile of isophotal pixels along major axis [pix]. \\
CLASS\_STAR\dotfill\				&	SE star/galaxy classification. \\
CXX\_IMAGE\dotfill\					&	SE ellipse parameter for isophotal pixels in Cartesian system [pix$^{-2}$]. \\
CXY\_IMAGE\dotfill\					&	SE ellipse parameter for isophotal pixels in Cartesian system [pix$^{-2}$]. \\
CYY\_IMAGE\dotfill\					&	SE ellipse parameter for isophotal pixels in Cartesian system [pix$^{-2}$]. \\
DEC\dotfill\						&	SE declination of object barycenter [J2000]. \\
FWHM\_IMAGE\dotfill\				&	SE FWHM assuming isophotal pixels have Gaussian light distribution [pix]. \\
ISOAREA\_IMAGE\dotfill\				&	SE isophotal area of pixels that satisfy analysis threshold (same as our detection threshold) [pix$^{2}$]. \\
RA\dotfill\							&	SE right ascension of object barycenter [J2000]. \\
THETA[ERR]\_IMAGE\dotfill\			&	SE ellipse position angle of isophotal pixels (CCW w/r/t x-axis) [deg]. \\
THETAWIN[ERR]\_IMAGE\dotfill\		&	SE windowed (Gaussian) ellipse position angle of isophotal pixels (CCW w/r/t x-axis) [deg]. \\
X\_IMAGE\dotfill\					&	SE object barycenter of isophotal pixels along x axis [pix]. \\
XWIN\_IMAGE\dotfill\				&	SE windowed (Gaussian) barycenter of isophotal pixels along x axis [pix]. \\
Y\_IMAGE\dotfill\					&	SE object barycenter of isophotal pixels along y axis [pix]. \\
YWIN\_IMAGE\dotfill\				&	SE windowed (Gaussian) barycenter of isophotal pixels along y axis [pix]. \\
\hline \multicolumn{2}{c}{\bf Ancillary Data} \\ \hline
ADAMI\_B[ERR]\dotfill\				&	\cite{Adami2006CAT} CFHT CFHK12 B-band magnitude (corrected for Gal. extinction) [instr. Vega]. \\
ADAMI\_V[ERR]\dotfill\				&	\cite{Adami2006CAT} CFHT Mould V-band magnitude (corrected for Gal. extinction) [instr. Vega]. \\
ADAMI\_R[ERR]\dotfill\				&	\cite{Adami2006CAT} CFHT Mould R-band magnitude (corrected for Gal. extinction) [instr. Vega]. \\
ADAMI\_I[ERR]\dotfill\				&	\cite{Adami2006CAT} CFHT Mould I-band magnitude (corrected for Gal. extinction) [instr. Vega]. \\
E\_BV\dotfill\						&	Color excess from \cite{Schlegel1998} reddening maps [mag]. \\
SDSS\_RA\dotfill\					&	SDSS DR6 right ascension [J2000]. \\
SDSS\_DEC\dotfill\					&	SDSS DR6 declination [J2000]. \\
SDSS\_PETROMAG\_U[ERR]\dotfill\	&	SDSS DR6 $u$-band Petrosian magnitude (not corrected for Gal. extinction) [asinh mag]. \\
SDSS\_PETROMAG\_G[ERR]\dotfill\	&	SDSS DR6 $g$-band Petrosian magnitude (not corrected for Gal. extinction) [asinh mag]. \\
SDSS\_PETROMAG\_R[ERR]\dotfill\	&	SDSS DR6 $r$-band Petrosian magnitude (not corrected for Gal. extinction) [asinh mag]. \\
SDSS\_PETROMAG\_I[ERR]\dotfill\		&	SDSS DR6 $i$-band Petrosian magnitude (not corrected for Gal. extinction) [asinh mag]. \\
SDSS\_PETROMAG\_Z[ERR]\dotfill\		&	SDSS DR6 $z$-band Petrosian magnitude (not corrected for Gal. extinction) [asinh mag]. \\
\hline
\end{tabular}
\label{CatalogParams}
\end{table}
\end{document}

%% file: definitions.tex
\newcommand{\simgt}{\lower 2pt \hbox{$\, \buildrel {\scriptstyle >}\over {\scriptstyle\sim}\,$}}
\newcommand{\simlt}{\lower 2pt \hbox{$\, \buildrel {\scriptstyle <}\over {\scriptstyle\sim}\,$}}

\newcommand{\galfit}{\textsc{galfit}}
\newcommand{\SExtractor}{\textsc{SExtractor}}
\newcommand{\SCAMP}{\textsc{Scamp}}
\def\Reff{\ifmmode{R_\mathrm{eff}}\else{$R_\mathrm{eff}$}\fi}
\def\mueff{\ifmmode{\mu_\mathrm{eff}}\else{$\mu_\mathrm{eff}$ }\fi}
\def\nser{\ifmmode{n_\mathrm{Ser}}\else{$n_\mathrm{Ser}$}\fi}
\def\muobs{\ifmmode{\mu_\mathrm{obs}}\else{$\mu_\mathrm{obs}$ }\fi}

\def\meanmueffSE{\ensuremath{\langle\mu\rangle_\mathrm{e}^\mathrm{SE}}}


%% file: tab1.tex
\begin{deluxetable}{ccccccc}
\tablecolumns{7}
\tabletypesize{\scriptsize}
\tablecaption{Fields Observed for the HST-ACS Coma Treasury Program \label{field_table}}
\tablewidth{0pt}
\tablenum{1}
\tablehead{
\colhead{Field} & \colhead{R.A.} & \colhead{Dec} & \colhead{Orient} & \colhead{Dither} & \multicolumn{2}{c}{\underline{Exposure (sec)}}		\\
\colhead{(visit no.)} & \colhead{(J2000)} & \colhead{(J2000)} & \colhead{(deg)}  & \colhead{Positions} & \colhead{F814W} & \colhead{F475W} 	\\
\colhead{(1)} & \colhead{(2)} & \colhead{(3)} & \colhead{(4)}  & \colhead{(5)} & \colhead{(6)} & \colhead{(7)}}
\startdata
01 & 13 00 45.90 & 28 04 54.0 &	282 & 4	& 1400 & 2677 \\
02 & 13 00 30.30 & 28 04 54.0 &	282 & 4	& 1400 & 2677 \\
03 & 13 00 14.70 & 28 04 54.0 & 282 & 2 & 700 &	1397 \\
08 & 13 00 42.30 & 28 01 43.0 & 282 & 4 & 1400 & 2677 \\
09 & 13 00 26.70 & 28 01 43.0 & 282 & 4	& 1400 & 2677 \\      
\\
10 & 13 00 11.10 & 28 01 43.0 & 282 & 4 & 1400 & 2677 \\      
12 & 12 59 39.90 & 28 01 43.0 & 282 & 3 & 1050 & 2012 \\     
13 & 12 59 24.30 & 28 01 43.0 & 282 & 2	& 700 & 1372 \\     
14 & 12 59 08.70 & 28 01 43.0 & 282 & 3	& 1050 & 2037 \\
15 & 13 00 38.70 & 27 58 32.0 &	282 & 4	& 1400 & 2677 \\ 
\\
16 & 13 00 23.10 & 27 58 32.0 & 282 & 4 & 1400 & 2677 \\
18 & 12 59 51.90 & 27 58 32.0 &	282 & 4	& 1400 & 2677 \\
19 & 12 59 36.30 & 27 58 32.0 &	282 & 4	& 1400 & 2677 \\
22 & 13 00 35.10 & 27 55 21.0 &	282 & 4	& 1400 & 2560 \\
23 & 13 00 19.50 & 27 55 21.0 &	282 & 4	& 1400 & 2677 \\
\\
24 & 13 00 03.90 & 27 55 21.0 &	282 & 4	& 1400 & 2560 \\
25 & 12 59 48.30 & 27 55 21.0 &	282 & 4	& 1400 & 2560 \\
33 & 12 59 29.10 & 27 52 10.0 &	282 & 4	& 1400 & 2677 \\
45 & 12 58 21.58 & 27 27 40.7 &	318 & 4	& 1400 & 2677 \\
46 & 12 58 34.00 & 27 22 58.8 &	280 & 4	& 1400 & 2677 \\
\\
59 & 12 58 31.15 & 27 11 58.5 &	270.05 & 4 & 1400 & 2677 \\
63 & 12 56 29.80 & 27 13 32.8 &	265 & 4	& 1400 & 2512 \\
75 & 12 58 47.75 & 27 46 12.7 &	280 & 4	& 1400 & 2677 \\
78 & 12 57 10.80 & 27 24 18.0 &	314.52 & 4 & 1400 & 2677 \\
90 & 12 57 04.22 & 27 31 34.5 &	299.10 & 4 & 1400 & 2677 \\
\enddata
\end{deluxetable}

%% file: tab2.tex
\begin{deluxetable}{cccccccccc}
\tablecolumns{10}
\tabletypesize{\scriptsize}
\tablecaption{Bright Galaxies Subtracted from ACS Images \label{subtract_table}}
\tablewidth{0pt}
\tablenum{2}
\tablehead{
\colhead{Field} & \colhead{Galaxy Name} & \colhead{R.A.} & \colhead{Dec} & \multicolumn{2}{c}{\underline{  F814W Magnitude (AB)  }} & \multicolumn{2}{c}{\underline{  F475W Magnitude (AB)  }} & \colhead{$r_{min}$} & \colhead{$r_{max}$} \\
\colhead{(visit no.)} & \colhead{(GMP ID)} & \colhead{(J2000)} & \colhead{(J2000)} & \colhead{{\tt MAG\_AUTO}}  & \colhead{{\tt MAG\_ISO}} & \colhead{{\tt MAG\_AUTO}}  & \colhead{{\tt MAG\_ISO}} & \colhead{(arcsec)} & \colhead{(arcsec)} \\
\colhead{(1)} & \colhead{(2)} & \colhead{(3)} & \colhead{(4)} & \colhead{(5)} & \colhead{(6)} & \colhead{(7)} & \colhead{(8)} & \colhead{(9)} & \colhead{(10)}}
\startdata
01 & 2440 & 195.20269 & 28.09075 & 13.55 & 13.51 &  14.90 & 14.87 & 7.5	& 20.0 \\
03 & 2861 & 195.05362 & 28.07549 & 14.66 & 14.63 &  15.94 & 15.91 & 9.0 & 30.0 \\
08 & 2417 & 195.21444 & 28.04302 & 13.45 & 13.41 &  14.81 & 14.78 & 4.0 & 20.0 \\
08 & 2551 & 195.16151 & 28.01454 & 14.39 & 14.38 &  15.63 & 15.62 & 2.5 & 12.0 \\
09 & 2727 & 195.09238 & 28.04703 & 14.28 & 14.26 &  15.54 & 15.53 & 4.5 & 15.0 \\
\\
10 & 2839 & 195.06144 & 28.04130 & 14.34 & 14.31 &  15.63 & 15.60 & 6.0 & 15.0 \\
10 & 2940 & 195.02665 & 28.00443 & 14.77 & 14.81 &  16.04 & 16.08 & 5.3 & 15.0 \\
13 & 3390 & 194.88106 & 28.04656 & 14.34 & 14.34 &  15.63 & 15.62 & 3.8 & 20.0 \\
15 & 2535 & 195.17016 & 27.99661 & 14.33 & 14.31 &  15.63 & 15.62 & 3.0 & 25.0 \\ 
15 & 2516 & 195.17819 & 27.97150 & 13.74 & 13.72 &  15.10 & 15.08 & 5.0 & 20.0 \\
\\
15 & 2510 & 195.17847 & 27.96304 & 14.52 & 14.51 &  15.82 & 15.82 & 5.5 & 15.0 \\
16 & 2815 & 195.06890 & 27.96754 & 14.58 & 14.56 &  15.75 & 15.74 & 9.0 & 15.0 \\
16 & 2654 & 195.11653 & 27.95599 & 14.70 & 14.69 &  16.00 & 15.99 & 4.5 & 22.0 \\
16 & 2651 & 195.11821 & 27.97240 & 14.65 & 14.65 &  15.94 & 15.94 & 1.5 & 20.0 \\
18 & 3170 & 194.94493 & 27.97389 & 14.21 & 14.20 &  15.53 & 15.53 & 4.2 & 25.0 \\
\\
19 & 3367 & 194.88655 & 27.98362 & 13.88 & 13.86 &  15.18 & 15.17 & 6.0 & 35.0 \\
19 & 3329 & 194.89874 & 27.95927 & 13.38 & 13.36 &  14.76 & 14.74 & 0.0 & 45.0 \\
19 & 3414 & 194.87482 & 27.95646 & 14.15 & 14.13 &  15.51 & 15.50 & 9.6 & 22.0 \\
19 & 3213 & 194.93219 & 27.99469 & 14.71 & 14.68 &  16.02 & 15.99 & 3.05 & 20.0 \\
22 & 2541 & 195.16564 & 27.92396 & 13.80 & 13.78 &  15.12 & 15.11 & 3.5 & 25.0 \\
\\
25 & 3201 & 194.93503 & 27.91246 & 13.89 & 13.87 &  15.21 & 15.19 & 4.5 & 20.0 \\
25 & 3222 & 194.92626 & 27.92477 & 14.93 & 14.89 &  16.20 & 16.16 & 4.5 & 10.0 \\
33 & 3423 & 194.87252 & 27.85013 & 14.06 & 14.02 &  15.42 & 15.38 & 8.0 & 25.0 \\
33 & 3400 & 194.87843 & 27.88418 & 13.78 & 13.73 &  15.09 & 15.05 & 7.0 & 35.0 \\
45 & 4206 & 194.63355 & 27.45635 & 14.79 & 14.77 &  16.07 & 16.05 & 5.5 & 31.0 \\
\\
46 & 4192 & 194.63806 & 27.36437 & 14.65 & 14.65 &  15.87 & 15.89 & 8.0 & 33.0 \\
75 & 3958 & 194.71707 & 27.78504 & 14.26 & 14.23 &  15.54 & 15.51 & 5.0 & 34.0 \\
78 & 5038 & 194.29484 & 27.40483 & 14.68 & 14.67 &  15.88 & 15.89 & 4.2 & 42.0 \\
\enddata
\end{deluxetable}

%% file: tab3.tex
\begin{deluxetable}{cclllcccccccc}
\tablecolumns{13}
\tabletypesize{\scriptsize}
\tablecaption{Simulated 80\% Completeness Limits Derived for a Range of Structural Parameters \label{Tab:80percent}}
\tablewidth{0pt}
\tablenum{3}
\tablehead{
\colhead{} & \multicolumn{6}{c}{\underline{F814W Logarithm(effective radius [arcsec])}} & \multicolumn{6}{c}{\underline{F475W Logarithm(effective radius [arcsec])}} \\
\colhead{Model} & \colhead{Pt Src} & \colhead{-1.08} & \colhead{-0.74} & \colhead{-0.39} & \colhead{-0.05} & \colhead{0.30} & \colhead{Pt Src} & \colhead{-1.08} & \colhead{-0.74} & \colhead{-0.39} & \colhead{-0.05} & \colhead{0.30} \\
\colhead{(1)} & \colhead{(2)} & \colhead{(3)} & \colhead{(4)} & \colhead{(5)} & \colhead{(6)} & \colhead{(7)} & \colhead{(8)} & \colhead{(9)} & \colhead{(10)} & \colhead{(11)} & \colhead{(12)} & \colhead{(13)}} \\
\startdata
FC             & 26.8    & 26.5      & 26.0  & 24.9  & 23.5  & 21.5 & 27.8  & 27.5      & 27.0  & 25.5  & 24.5  & 22.0   \\       
FE             & 26.8 	 & 26.8     & 26.0  & 25.3 & 24.0  & 22.0 & 27.8  & 27.5      & 27.0  & 26.0  & 25.0  & 23.0   \\       
PC             & 26.8 	 & 26.8     & 26.3 & 25.5  & 24.5  & 23.0 & 27.8  & 27.5      & 27.0  & 26.3  & 25.5  & 23.9   \\       
PE             & 26.8 	 & 26.8      & 26.1  & 25.5  & 24.7  & 23.5 & 27.8  & 27.7      & 27.2  & 26.5  & 25.5  & 24.5   \\
\enddata
\tablecomments{F/P (flat/peak) models have S\'ersic indices smaller/larger than \nser$=$2.25, and C/E (circular/elliptical) models have ellipticity smaller/larger than $e$=0.4; logarithmic effective radii are given at the center of each bin, which are equally-spaced in dex over the range \Reff=0.5-60 pixels.}
\end{deluxetable}

%% file: tab4.tex
\begin{deluxetable}{ccccccccccccccc}
\tablecolumns{15}
\tabletypesize{\scriptsize}
\tablecaption{K-corrections for Galaxy SEDs at the Distance of Coma \label{kcorr}}
\tablewidth{0pt}
\tablenum{4}
\tablehead{
\colhead{} & \multicolumn{3}{c}{\underline{  Ellipticals   (E)  }} & \colhead{S0} & \multicolumn{2}{c}{\underline{Sa}} & \colhead{Sb} & \colhead{Sbc} & \multicolumn{2}{c}{\underline{Sc}} & \colhead{Scd} & \colhead{Im} & \colhead{SB4} & \colhead{SB1}\\
\colhead{Band} & \colhead{(K96)} & \colhead{(P97)} & \colhead{(C80)} & \colhead{(K96)} & \colhead{(K96)} & \colhead{(P97)} & \colhead{(K96)} & \colhead{(C80)} & \colhead{(K96)} & \colhead{(P97)} & \colhead{(C80)} & \colhead{(C80)} & \colhead{(K96)} & \colhead{(K96)}	\\
\colhead{(1)} & \colhead{(2)} & \colhead{(3)} & \colhead{(4)}  & \colhead{(5)} & \colhead{(6)} & \colhead{(7)} & \colhead{(8)} & \colhead{(9)} & \colhead{(10)} & \colhead{(11)} & \colhead{(12)} & \colhead{(13)} & \colhead{(14)} & \colhead{(15)}}	\\
\startdata
F475W & 0.09 & 0.08 & 0.08 & 0.08 & 0.05 & 0.06 & 0.05 & 0.03 & 0.04 & 0.03 & 0.02 & 0.04 & 0.05 & 0.10		\\
F814W & 0.02 & 0.01 & 0.01 & 0.00 & 0.01 & 0.01 & 0.01 & 0.02 & \nodata & 0.00 & 0.00 & 0.00 & 0.02 & 0.00	\\
\enddata
\end{deluxetable}